%% file: MITL-to-CLTL.tex
\documentclass[copyright,creativecommons]{eptcs}
\providecommand{\event}{GandALF 2013} 

\usepackage{amsmath,amssymb,amsthm}
\usepackage{mathrsfs}
\usepackage{savesym}
\savesymbol{amalg}
\usepackage{mathabx}
\restoresymbol{ABXF}{amalg}
\usepackage{bm}
\usepackage{multirow}
\usepackage[textwidth=2.0cm,textsize=scriptsize]{todonotes}
\usetikzlibrary{arrows}

\newtheorem{theorem}{Theorem} 
\newtheorem{lemma}{Lemma} 
\newtheorem{corollary}{Corollary}

\input{macros}

\newif\iflong
\longfalse

\newif\ifremarks
\remarksfalse

\newif\iffinalversion

\begin{document}

\title{Deciding the Satisfiability of 
MITL Specifications\thanks{This research was supported by the Programme IDEAS-ERC, Project 227977-SMScom.}}

\author{
Marcello M. Bersani$^\ast$ \qquad Matteo Rossi$^\ast$ \qquad Pierluigi {San Pietro}$^{\ast,+}$
\institute{${}^\ast$ Dipartimento di Elettronica Informazione e Bioingegneria, Politecnico di Milano, Milano, Italy \\
${}^+$CNR IEIIT-MI, Milano, Italy}
\email{\{marcellomaria.bersani,matteo.rossi,pierluigi.sanpietro\}@polimi.it}
}

\def\titlerunning{Deciding the Satisfiability of 
MITL Specifications}
\def\authorrunning{M. M. Bersani, M. Rossi, \& P. {San Pietro}}


\maketitle
%

\input{abstract}

\input{section-intro}

\input{section-languages}

\input{section-reduction}

\input{section-generalization}

\input{section-conclusions}

\bibliographystyle{eptcs}
\bibliography{bibliografia}

%
%


\end{document}

%% file: macros.tex
\newcommand{\D}{\mathcal{D}}

\newcommand{\U}{\mathbf{U}}
\newcommand{\Snc}{\mathbf{S}}
\newcommand{\T}{\mathbf{T}}
\newcommand{\R}{\mathbf{R}}

\newcommand{\Nat}{\mathbb{N}}

\newcommand{\Real}{\mathbb{R}}

\newcommand{\X}[1]{\mathbf{X}\left(#1\right)}
\newcommand{\Y}[1]{\mathbf{Y}\left(#1\right)}

\newcommand{\iFF}{\Leftrightarrow}

\newcommand{\G}[1]{\mathbf{G}\left(#1\right)}

\newcommand{\pair}[2]{(#1, #2)}

\newcommand{\aX}{\mathrm{X}}

\makeatletter
\def\Eqlfill@{\arrowfill@\Relbar\Relbar\Relbar}
\newcommand{\longmodels}[1][]{\,|\!\!\!\ext@arrow 0359\Eqlfill@{#1}}
\makeatother

\newcommand{\interval}[2]{\langle #1,#2 \rangle}

\newcommand{\set}[1]{\{ #1 \}}

\usepackage{mathtools}
\changenotsign
\usepackage[electronic]{ifsym}
\usepackage{stackrel}

\def\NOTRaisingEdge{{%
    \setbox0\hbox{\RaisingEdge}%
    \rlap{\hbox to \wd0{\hss/\hss}}\box0
}}

\def\NOTFallingEdge{{%
    \setbox0\hbox{\FallingEdge}%
    \rlap{\hbox to \wd0{\hss/\hss}}\box0
}}

\def\NOTShortPulseHigh{{%
    \setbox0\hbox{\ShortPulseHigh}%
    \rlap{\hbox to \wd0{\hss/\hss}}\box0
}}

\def\NOTShortPulseLow{{%
    \setbox0\hbox{\ShortPulseLow}%
    \rlap{\hbox to \wd0{\hss/\hss}}\box0
}}

  \DeclareFontFamily{U}  {MnSymbolA}{}
  \DeclareSymbolFont{MnSyA}         {U}  {MnSymbolA}{m}{n}
  \SetSymbolFont{MnSyA}       {bold}{U}  {MnSymbolA}{b}{n}
\DeclareFontShape{U}{MnSymbolA}{m}{n}{
    <-6>  MnSymbolA5
   <6-7>  MnSymbolA6
   <7-8>  MnSymbolA7
   <8-9>  MnSymbolA8
   <9-10> MnSymbolA9
  <10-12> MnSymbolA10
  <12->   MnSymbolA12}{}
\DeclareFontShape{U}{MnSymbolA}{b}{n}{
    <-6>  MnSymbolA-Bold5
   <6-7>  MnSymbolA-Bold6
   <7-8>  MnSymbolA-Bold7
   <8-9>  MnSymbolA-Bold8
   <9-10> MnSymbolA-Bold9
  <10-12> MnSymbolA-Bold10
  <12->   MnSymbolA-Bold12}{}
  \DeclareMathSymbol{\leftfilledspoon}{\mathrel}{MnSyA}{114}
  \DeclareMathSymbol{\leftspoon}{\mathrel}{MnSyA}{106}
  \DeclareMathSymbol{\upfilledspoon}{\mathrel}{MnSyA}{113}
  \DeclareMathSymbol{\upspoon}{\mathrel}{MnSyA}{105}
  \DeclareMathSymbol{\leftfree}{\mathrel}{MnSyA}{130}

  \DeclareFontFamily{U}  {MnSymbolB}{}
  \DeclareSymbolFont{MnSyB}         {U}  {MnSymbolB}{m}{n}
  \SetSymbolFont{MnSyB}       {bold}{U}  {MnSymbolB}{b}{n}
\DeclareFontShape{U}{MnSymbolB}{m}{n}{
    <-6>  MnSymbolB5
   <6-7>  MnSymbolB6
   <7-8>  MnSymbolB7
   <8-9>  MnSymbolB8
   <9-10> MnSymbolB9
  <10-12> MnSymbolB10
  <12->   MnSymbolB12}{}
\DeclareFontShape{U}{MnSymbolB}{b}{n}{
    <-6>  MnSymbolB-Bold5
   <6-7>  MnSymbolB-Bold6
   <7-8>  MnSymbolB-Bold7
   <8-9>  MnSymbolB-Bold8
   <9-10> MnSymbolB-Bold9
  <10-12> MnSymbolB-Bold10
  <12->   MnSymbolB-Bold12}{}
  \DeclareMathSymbol{\nleftfilledspoon}{\mathrel}{MnSyB}{114}

\newcommand{\rest}[1]{\ensuremath{\stackrel{\leftfree}{#1}}}
\newcommand{\fst}[1]{\ensuremath{\upfilledspoon_{#1}}}
\newcommand{\allup}[1]{\ensuremath{\stackrel{\leftfilledspoon}{#1}}}
\newcommand{\alldn}[1]{\ensuremath{\stackrel[\leftfilledspoon]{}{#1}}}

\newcommand{\Hg}[1]{\ensuremath{\textrm{\RaisingEdge}_{#1}}}

\newcommand{\Lw}[1]{\ensuremath{\textrm{\FallingEdge}_{#1}}}

\newcommand{\Sup}[1]{\ensuremath{\textrm{\ShortPulseHigh}_{#1}}}

\newcommand{\Sdn}[1]{\ensuremath{\textrm{\ShortPulseLow}_{#1}}}

\newcommand{\qtlSup}[1]{\ensuremath{s^{u}_{#1}}}

\newcommand{\qtlSdn}[1]{\ensuremath{s^{d}_{#1}}}

\DeclareRobustCommand{\shortf}[3] 
{\ensuremath{%
	 \ifthenelse{\not \equal{#3}{} }  {\stackrel{#2}{#1}_{#3}}  {\stackrel{#2}{#1}}
  }%
}%

\def\dush{\mathrel{\rotatebox[origin=c]{-90}{$\Lsh$}}}

\DeclareRobustCommand{\befDnowU}[2]{\shortf{\dush}{#1}{#2}}

\DeclareRobustCommand{\notbefDnowU}[2]{\shortf{\not \dush}{#1}{#2}}

\DeclareRobustCommand{\nowonD}[2]{\shortf{\drsh}{#1}{#2}}

\DeclareRobustCommand{\orig}{\ensuremath{O}}

\newcommand{\myitem}[1]{$\bullet \quad$ #1:~\hspace{6pt}~}
\newcommand{\sig}{\ensuremath{M}}

\newcommand{\rmap}[2]{\ensuremath{r_{#1}(#2)}}
\newcommand{\rmapinv}[2]{\ensuremath{r^{-1}_{#1}(#2)}}

\DeclareRobustCommand{\LogOp}[3] 
{\ensuremath{%
	 \ifthenelse{\not \equal{#2}{}}  {\mathbf{#1}_{#2}}  {\mathbf{#1}}
	 \ifthenelse{\not \equal{#3}{}}  {\!\left({#3}\right)} {}
  }%
}%

\DeclareRobustCommand{\LogOpPast}[3] 
{\ensuremath{%
	 \ifthenelse{\not \equal{#2}{}}  {\overleftarrow{\mathbf{#1}}_{#2}}  {\mathbf{#1}}
	 \ifthenelse{\not \equal{#3}{}}  {\!\left({#3}\right)} {}
  }%
}%

\DeclareRobustCommand{\Ultl}[1]{\LogOp{U}{#1}{}}
\DeclareRobustCommand{\Sltl}[1]{\LogOp{S}{#1}{}}

\DeclareRobustCommand{\Xltl}[1]{\LogOp{X}{}{#1}}
\DeclareRobustCommand{\Yltl}[1]{\LogOp{Y}{}{#1}}
\DeclareRobustCommand{\Gltl}[2]{\LogOp{G}{#1}{#2}}
\DeclareRobustCommand{\Fltl}[2]{\LogOp{F}{#1}{#2}}
\DeclareRobustCommand{\Pltl}[2]{\LogOp{P}{#1}{#2}}

\DeclareRobustCommand{\LogOpInfix}[4] 
{\ensuremath{%
	 \ifthenelse{\not \equal{#3}{}}  {#3}  {}
	 \ifthenelse{\not \equal{#2}{}}  {\mathbf{#1}_{#2}}  {\mathbf{#1}}
	 \ifthenelse{\not \equal{#4}{}}  {#4} {}
  }%
}%

\DeclareRobustCommand{\Uinfix}[3]{\LogOpInfix{U}{#1}{#2}{#3}}
\DeclareRobustCommand{\Sinfix}[3]{\LogOpInfix{S}{#1}{#2}{#3}}

%% file: abstract.tex
\begin{abstract}
In this paper we present a satisfiability-preserving reduction from MITL interpreted over finitely-variable continuous behaviors to Constraint LTL over clocks, a variant of CLTL that is decidable, and for which an SMT-based bounded satisfiability checker is available.
The result is a new complete and effective decision procedure for MITL.
Although decision procedures for MITL already exist, the automata-based techniques they employ appear to be very difficult to realize in practice, and, to the best of our knowledge, no implementation currently exists for them.
A prototype tool for MITL based on the encoding presented here has, instead, been implemented and is publicly available. 
\end{abstract} 

%% file: section-intro.tex
\section{Introduction}
\label{section-intro}

Computer systems are inherently discrete-time objects, but their application to control and monitoring of real-time systems often requires to deal with time-continuous external signals and variables, such as position, speed and acceleration or temperature and pressure. Hence, many continuous-time models have been developed for verification and validation of such systems, e.g., Timed Automata~\cite{Alur&Dill94}, or continuous-time temporal logics, such as MITL (Metric Interval Temporal Logic)~\cite{AFH96}. 

In general, the role of temporal logics in verification and validation is two-fold. 
First, temporal logic allows abstract, concise and convenient expression of required properties of a system.
Linear Temporal Logic (LTL) is often used with this goal in the verification of finite-state models, e.g., in model checking~\cite{BK08}. 
Second, temporal logic allows a descriptive approach to specification and modeling (see, e.g.,~\cite{MS94,FMMR12}).  
A descriptive model is based on axioms, written in some (temporal) logic, defining  a system by means of its general properties, rather than by an
operational model based on some kind of machine (e.g., a Timed Automaton) behaving in the desired way.
In this case, verification typically consists of satisfiability checking of the conjunction of the model and of the (negation of) its desired properties. 
An example of the latter approach is Bounded Satisfiability Checking (BSC) \cite{PMS12}, where Metric Temporal Logic (MTL) specifications on {\em discrete} time and properties
are translated into Boolean logic, in an approach similar to Bounded Model Checking of LTL properties of finite-state machines.

In general, verification of continuous-time temporal logics is not as well sopported  as for discrete-time models. 
Uppaal~\cite{BY04} is the de-facto standard tool for verification of Timed Automata. 
However, Uppaal does not support continuous-time temporal logics: 
not only satisfiability checking is not available in Uppaal, but even the formalization of system properties in temporal logic is not allowed, aside from rather simple invariants and reachability properties.  
Rather, non-trivial properties to be verified on an operational model must be expressed as other Timed Automata, i.e., 
at a lower level of abstraction. 
Indeed, there have been a few proposals for verifying continuous-time logics \cite{MNP06}, but they do not appear to be actually implementable, and, to the best of our knowledge,  in fact they have never been implemented.

This paper proposes a new technique, based on generalizing BSC to MITL, by reducing satisfiability of MITL to satisfiability of {\em Constraint LTL} {\em over clocks} (CLTL-oc), 
a new decidable variant of CLTL~\cite{DD07}.
In particular, a MITL formula may be encoded into an equisatisfiable CLTL-oc formula, which can then be solved through the same techniques of~\cite{BFMPRS10,BFRS11,BFMPRS12}. 
The latter approach generalizes BSC to CLTL, generating an encoding suitable for verification with standard Satisfiability Modulo Theories (SMT) solvers such as Z3~\cite{z3}.
This new technique has been implemented in an open-source prototype tool \cite{qtlsolver}.

Although MITL is known to be decidable over unrestricted behaviors \cite{HR04}, we focus on so-called finitely-variable models, i.e. such that in every bounded time interval there can only be a finite number of changes.
This is a very common requirement for continuous-time models, which only rules out pathological behaviors (e.g., Zeno
\cite{FMMR12}) which do not have much practical interest.
To define the encoding, we start by focusing on models in which intervals are closed on the left end and open on the right end.
This restriction is later lifted to consider general, finitely-variable, signals.

The paper is organized as follows: Sect.~\ref{section-languages} defines MITL and CLTL-oc, Sect.~\ref{section-reduction} defines a reduction from MITL to CLTL-oc,
based on the restriction that intervals are closed to the left and open to the right;
Sect.~\ref{section-generalization} generalizes the translation to intervals of any kind,  also discussing the extension to include past operators.
Sect.~\ref{section-conclusions} concludes, discussing applications to other logics and presenting a prototype tool.

%% file: section-languages.tex
\section{Languages}\label{section-languages}

Let $AP$ be a finite set of atomic propositions.
The syntax of (well formed) formulae of MITL is defined as follows, with $p\in AP$ and
$I$ an interval of the form $\interval{a}{b}$ or 
$\interval{a}{+\infty}$, with
$a,b \in \Nat$ constants, 
$a< b$:
\begin{equation*}
  \phi :=
  \begin{gathered}
    p \mid \phi \wedge \phi \mid \neg \phi \mid \phi\U_I\phi
  \end{gathered}
\end{equation*}
%
%
\begin{table}[bt]
\begin{equation*}
\begin{split}
\sig, t \models p                &\iFF  p\in \sig(t)  \qquad p\in AP\\
\sig, t \models \neg \phi        &\iFF  \sig,t \not\models \phi \\
\sig, t \models \phi \wedge \psi &\iFF  \sig, t \models \phi \text{ and } \sig, t \models \psi\\
\sig, t \models \phi\U_I\psi     &\iFF \exists t'\in t+I: \sig, t' \models \psi \text{ and } \sig,t'' \models \phi \ \forall t'' \in (t,t') \\
\end{split}
\end{equation*}
\caption{Semantics of MITL.}
\label{tab:MITLsemantics}
\end{table}
%
The semantics of MITL is defined in Table~\ref{tab:MITLsemantics} with respect to \emph{signals}. 
A signal is a function $\sig:\Real_+ \to 2^{AP}$, with $\Real_+$ the set of nonnegative reals.
A MITL formula $\phi$ is \emph{satisfiable} if there exists
a signal $\sig$ such that $\sig, 0 \models \phi$ (in this case, $\sig$ is called a \emph{model} of $\phi$).
%
%
%
%
The \emph{globally} $\mathbf{G}_I$ and \emph{eventually} $\mathbf{F}_I$ operators can be defined by the usual abbreviations: 
$\mathbf{F}_I \phi = \top \U_I \phi$ and $\mathbf{G}_I \phi = \neg \mathbf{F}_I (\neg \phi)$.
{\em Constraint LTL} (CLTL~\cite{DD07,BFRS11}) is used in Sect.~\ref{section-reduction} to solve the satisfiability problem of MITL.
CLTL formulae are defined with respect to a finite set $V$ of variables and a {\em constraint system} $\D$, which is a pair
$(D, \mathcal{R})$ with $D$ being a specific domain of interpretation for variables and
constants and
$\mathcal{R}$ being a family of relations on $D$, such that 
the set $AP$ of atomic propositions coincides with set $\mathcal{R}_0$ of 0-ary relations.
An {\em atomic constraint} is a term of the form
$R(x_1,\dots,x_n)$, where $R$
is an $n$-ary relation of $\mathcal{R}$ on domain $D$ and $x_1, \dots, x_n$ are variables.
A \emph{valuation} is a mapping $v: V \to D$, i.e., an assignment of a value in $D$
to each variable.
A constraint is {\em satisfied} by $v$, written $v \models_\D R(x_1,\dots,x_n)$, if $( v(x_1),\dots,v(x_n) ) \in R$. 
Given a variable $x \in V$ over domain $D$, \emph{temporal terms} are defined by the syntax:
$   \alpha := c \mid x \mid \aX \alpha$,
where $c$ is a constant in $D$ and $x$ denotes a variable over $D$.
Operator $\aX$ is very similar to $\mathbf{X}$, but it only applies to temporal terms,  
with the meaning that $\aX \alpha$ is the \emph{value}
of temporal term $\alpha$ in the next time instant.
Well-formed CLTL formulae are defined as follows:
\begin{equation*}
  \phi :=
    R(\alpha_1, \dots, \alpha_n) \mid \phi \wedge \phi \mid \neg \phi \mid
   \X{\phi} \mid \Y{\phi} 
\mid \phi\U\phi \mid \phi\Snc\phi
\end{equation*}
where $\alpha_i$'s are temporal terms, $R \in \mathcal{R}$, $\mathbf{X}$, $\mathbf{Y}$, $\U$ and $\Snc$ are the
usual ``next'', ``previous'', ``until'' and ``since'' operators of LTL, with the same meaning.
The dual operators ``release'' $\R$, and ``trigger'' $\T$ 
may be defined as usual, i.e., $\phi\R \psi$ is $\neg(\neg \phi \U \neg\psi)$ and 
$\phi\T \psi$ is $\neg(\neg \phi \Snc \neg\psi)$.
The semantics of CLTL formulae is defined with respect to
a strict linear order representing time $\pair{\Nat}{<}$.
Truth values of propositions in $AP$, and values of variables belonging
to $V$ are defined by a pair
$\pair{\pi}{\sigma}$ where $\sigma : \Nat \times V \to D$ is a function
which defines the value of variables at each position in $\Nat$ and
$\pi :\Nat \to \wp(AP)$ is a function associating a subset of the set of
propositions with each element of $\Nat$.
The value of terms is defined with respect to $\sigma$ as follows:
\[
\sigma(i, \alpha) = \sigma(i+|\alpha|, x_{\alpha})
\]
where $x_\alpha$ is the variable in $V$ occurring in term $\alpha$ and $|\alpha|$ is the \emph{depth} of a temporal term, namely 
the total amount of temporal shift needed in evaluating $\alpha$: 
$|x| = 0$ when $x$ is a variable, and $|\aX\alpha| = |\alpha| + 1$. 
The semantics of a CLTL formula $\phi$ at instant $i\geq 0$
over a linear structure $\pair{\pi}{\sigma}$ is recursively defined as in Table~\ref{tab:CLTLsemantics},
where $R\in\mathcal{R}\setminus\mathcal{R}_0$.
\begin{table}[tb]
\begin{equation*}
\begin{aligned}
\pair{\pi}{\sigma}, i \models p  &\iFF  p \in \pi(i) \text{ for } p \in AP \\
\pair{\pi}{\sigma}, i \models R(\alpha_1, \dots, \alpha_n) &\iFF
  (\sigma(i+|\alpha_1|, x_{\alpha_1}), \dots, \sigma(i+|\alpha_n|, x_{\alpha_n}) ) \in R\\
\pair{\pi}{\sigma}, i \models \neg \phi &\iFF  \pair{\pi}{\sigma},i \not\models \phi \\
\pair{\pi}{\sigma}, i \models \phi \wedge \psi &\iFF  \pair{\pi}{\sigma}, i \models \phi
\, \text{and} \, \pair{\pi}{\sigma}, i \models \psi\\
\pair{\pi}{\sigma}, i \models \X \phi &\iFF \pair{\pi}{\sigma},i+1 \models \phi \\
\pair{\pi}{\sigma}, i \models \Y \phi &\iFF \pair{\pi}{\sigma},i-1 \models
\phi \wedge i>0\\
\pair{\pi}{\sigma}, i \models \phi\U\psi &\iFF
\exists \, j\geq i: \pair{\pi}{\sigma},j \models \psi \ \wedge \pair{\pi}{\sigma},n \models \phi \ \forall \ i\leq n < j \\
\pair{\pi}{\sigma}, i \models \phi\Snc\psi &\iFF
\exists \, 0\leq j \leq i: \pair{\pi}{\sigma},j \models \psi \, \wedge  \pair{\pi}{\sigma},n \models \phi \ \forall \ j < n \leq i \\ 
\end{aligned}
\end{equation*}
\caption{Semantics of CLTL.}\label{tab:CLTLsemantics}
\end{table}
A formula $\phi \in$ CLTL is \emph{satisfiable} if there exists 
a pair $\pair{\pi}{\sigma}$ such that $\pair{\pi}{\sigma},0 \models
\phi$. 


In this paper, we consider a variant of CLTL, where arithmetic variables are evaluated as \emph{clocks} and set $\mathcal{R}$ is $\set{<,=}$.
A clock ``measures'' the time elapsed since the last time the clock was ``reset'' (i.e., the variable was equal to 0).
By definition, in CLTL-oc each $i \in \Nat$ is associated with a ``time delay'' $\delta(i)$, where $\delta(i) > 0$ for all $i$, which corresponds to the ``time elapsed'' between $i$ and the next state $i+1$.
More precisely, for all clocks $x\in V$, $\sigma(i+1, x) = \sigma(i,x) + \delta(i)$, unless it is ``reset'' (i.e., $\sigma(i+1,x) = 0$).

%% file: section-reduction.tex
\section{Reduction of MITL to CLTL-over-clocks}
\label{section-reduction}

This section devises a reduction from MITL to CLTL-oc.
The inherent bounded variability of metric operators in MITL allows a translation  of a MITL formula $\phi$ into  a CLTL-oc formula with a bounded number of variables, depending on the subformulae of $\phi$.

As in \cite{MNP06,Souza&Tabareau04}, it is actually convenient to introduce the operators $\U_{(0,+\infty)}$ and $\mathbf{F}_I$ as primitive, and instead derive the metric until $\U_I$, as shown by 
the following result.

\begin{lemma}\label{lemma-metric-until}
Let $\sig$ be a signal. Then, for any $t\geq 0$, 
$$
\begin{aligned}
(1) \ \ \sig, t \models \phi \U_{[ a,b \rangle} \psi &\iFF \sig, t \models \mathbf{G}_{[0,a)} (\phi \U_{(0,+\infty)} \psi) \wedge \mathbf{F}_{[ a,b \rangle} \psi \\
(2) \ \ \sig, t \models \phi \U_{( a,b \rangle} \psi &\iFF \sig, t \models \mathbf{G}_{[0,a]} (\phi \U_{(0,+\infty)} \psi) \wedge \mathbf{F}_{( a,b \rangle} \psi \\
(3) \ \ \sig, t \models \phi \U_{\langle 0,b \rangle} \psi &\iFF \sig, t \models \phi \U_{\langle 0,+\infty)} \psi \wedge \mathbf{F}_{\langle 0,b \rangle} \psi \\
\end{aligned}
$$
When $b$ is $+\infty$, equivalences $(1),(2)$ can be simplified, respectively, in $\phi \U_{[ a,+\infty )} \psi \equiv \mathbf{G}_{[0,a)} (\phi \U_{(0,+\infty)} \psi)$ and $\phi \U_{( a,+\infty )} \psi \equiv \mathbf{G}_{[0,a]} (\phi \U_{(0,+\infty)} \psi)$.
\end{lemma}
The above equivalences  make it possible to base the CLTL-oc translation on the $\U_{( 0,+\infty )}$ and $\mathbf{F}_I$ operators, 
instead of $\U_I$, therefore confining metric issues only to the translation of $\mathbf{F}_I$, 
which is much simpler than the translation of $\U_I$.

Reducing MITL to CLTL-oc requires a way to represent models of MITL formulae, i.e., continuous signals over a finite set of atomic propositions, by means of CLTL-oc models where time is discrete.

Discrete positions in CLTL-oc models represent, for each subformula $\theta$ of $\phi$, the occurrence of an ``event''
at that point for the subformula.
An ``event'' is a change of truth value (``become true'' or ``become false'') of $\theta$. Hence, the signal is ``stable'' (i.e., there is no change) in the interval 
between two events: a continuous-time signal is hence partitioned by the above events into intervals.
Time progress between two discrete points is measured by CLTL variables behaving as clocks: 
for each subformula $\theta$ of $\phi$, there are two clocks 
$z^0_\theta,z^1_\theta$ 
measuring the time elapsed since the last ``become true'' and ``become false'' events, respectively (i.e., they are reset when the corresponding event occurs).
In case of subformulae of the form $\theta=\mathbf{F}_{\langle a,b \rangle} \phi$, also a finite set of auxiliary clocks is introduced, whose cardinality depends on the values of $a,b$, namely
$d=2\left\lceil\frac{b}{b-a}\right\rceil$ auxiliary clocks $x^j_\theta$  ($0\le j \le d-1$).
Therefore, a CLTL-oc model embeds, in every (discrete) position both  the information defining the truth value of all the subformulae occurring in $\phi$ and 
also the time progress between two consecutive events.
Then, every position in the CLTL-oc model captures the configuration of one of the intervals in which the MITL signals are partitioned by the events.
Therefore, our reduction defines, by means of CLTL-oc formulae, the semantics of every subformula of $\phi$.

We start by restricting the set of signals defining models of MITL formulae to signals where intervals are 
left-closed and right-open ({\em l.c.r.o.}), e.g.: \includegraphics[scale=0.5]{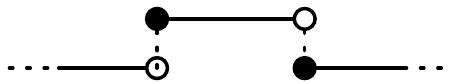}. 
We will lift this restriction later in the paper.
Hence, singularities (i.e., events being true in a single instant) cannot occur and may be ignored.
However, the semantics given here does not exclude \emph{a priori} Zeno behaviors \cite{FMMR12}: 
it admits signals corresponding to an infinite sequence of events accumulating to the left of a time instant, i.e.,
where events do not advance beyond that instant.
However, since these signals correspond to behaviors that are of little interest in practice, 
we restrict the set of models to non-Zeno signals,  i.e., to models of CLTL-oc formulae where  time diverges: $\sum_{i \in \Nat} \delta(i) = \infty$,
by enforcing a suitable CLTL-oc constraint.

Let $\sig$ be a signal, $\phi$ a MITL formula over $AP$ and $sub(\phi)$ the set of all subformulae occurring in $\phi$.
We write $\uparrow_\theta$ for the occurrence of an event making $\theta\in sub(\phi)$ become true. With abuse of notation
 we extend $\models$ as follows:
\[
\sig, t \models \uparrow_\theta \iFF 
\begin{gathered}
 \sig, t \models \theta \text{ and }
 \left(
\begin{gathered}
 \exists \varepsilon>0 \ \forall t' \in (t,t+\varepsilon) \ \sig, t' \models \theta \text{ and } \\
 t > 0\Rightarrow \exists \varepsilon>0 \ \forall t' \in (t-\varepsilon,t) \ \sig, t' \models \neg\theta 
\end{gathered}
\right)
\end{gathered}
\]
We define $\downarrow_\theta$ as an abbreviation for $\uparrow_{\neg\theta}$. 
These definitions impose that signals are defined over an infinite sequence of intervals of the form $[t_1, t_2)$ where $t_2 > t_1$.

Not all temporal operators preserve l.c.r.o. intervals. For example, let $\theta = \mathbf{F}_{\langle a,b )}\phi$ be a MITL formula and let $\phi$ hold on a l.c.r.o. signal;
then, the corresponding signal for $\theta$ (i.e., the signal including also the values for $\uparrow_\theta$),  is not l.c.r.o..
In fact, let $t>b$ be the first position such that $\sig,t \models \uparrow_\phi$.
If the signal for $\theta$ were l.c.r.o., then it should be $\sig, t-b \models \uparrow_\theta$, 
which is impossible because $\sig,t-b\models \mathbf{F}_{\langle a,b )}\phi \iFF \exists t'' \in t-b+\langle a, b) \ \sig,t'' \models \phi$ and $t'' < t$, but by hypothesis $\phi$ is false before $t$.
Nevertheless, the next result shows that
that Boolean connectives $\neg,\wedge$ and temporal operators $\U_{(0,+\infty)}$, $\mathbf{F}_{\langle a,b ]}$, $\mathbf{F}_{\langle a, +\infty)}$ and $\mathbf{F}_{\langle 0,b ]}$, 
do indeed preserve l.c.r.o. intervals.

We extend MITL models to any subformulae occurring in MITL formulae by
defining a mapping $\sig_\theta:\Real_+ \to \set{\emptyset, \theta}$ such that:
$$\theta \in \sig_\theta(t) \iFF \sig,t \models \theta.$$

\begin{lemma}\label{lemma-leftclosed-signals}
Let $\sig$ be a l.c.r.o. signal, let $\phi, \psi$ be two formulae occurring in $\sig$ and let $\theta$ be a formula 
$\neg\phi$, $\phi\wedge\psi$, $\U_{(0,+\infty)}(\phi,\psi), \mathbf{F}_{\langle a,b ]}(\phi), \mathbf{F}_{\langle a, +\infty)}(\phi), \mathbf{F}_{\langle 0,b ]}(\phi)$. 
Then, $\sig_\theta$ is a l.c.r.o. signal.
\end{lemma}

\noindent In what follows, $\mathbf{F}_{\langle a, +\infty)}$ is defined as primitive, instead of applying the known equivalence 
$\mathbf{F}_{[ a, +\infty)}\phi \equiv \top \U_{[ a,+\infty )} \phi \equiv \mathbf{G}_{[0,a)} (\phi \U_{(0,+\infty)} \psi)$,
as formula $\mathbf{G}_{[0,a)}\phi \equiv \neg\mathbf{F}_{[0,a)}\neg\phi$ 
violates the l.c.r.o. assumption. 

We now show how to build a CLTL-oc model $(\pi, \sigma)$ of $\phi$ from a signal $\sig$.
For each subformula $\theta \in sub(\phi)$ we introduce two \emph{clock variables} $z^0_\theta,z^1_\theta$
and one atomic proposition $\allup{\theta}$. We will ensure that $\allup{\theta}$  is true at a position whenever
$\theta$ is true in the interval corresponding to the position.
To ease understanding, in the rest we use $\alldn{\theta} = \neg \allup{\theta}$.
We also introduce two abbreviations, $\Hg{\theta}, \Lw{\theta}$ that play the role of \emph{event markers} (referred to as just ``events'' when the context is clear); 
more precisely, they denote, respectively, events $\uparrow_\theta$ and $\downarrow_\theta$, and are defined as follows: 
$$
\Hg{\xi} = \neg \Yltl{}({\allup{\xi}}) \wedge \allup{\xi}
\hspace{1.5cm}
\Lw{\xi} = \neg \Yltl{}({\alldn{\xi}}) \wedge \alldn{\xi}
$$
Note that, as $\neg \Y{\bullet}$ is true in the origin, no matter the argument, either $\Hg{\theta}$ or $\Lw{\theta}$ holds at 0.

For each $\theta=\mathbf{F}_{\langle a,b ]}\psi \in sub(\phi)$ we introduce $d=2\left\lceil\frac{b}{b-a}\right\rceil$ \emph{auxiliary clocks} $x^0_\theta, \dots x^{d}_\theta$.
The idea behind the above definitions is that at each occurrence of an event marker ($\Hg{\theta}$ or $\Lw{\theta}$), exactly one of the clocks $z^0_\theta,z^1_\theta$ is equal to 0; the clock, then, measures the time elapsed from the last opposite event. 
Instead, the auxiliary clocks associated with formulae $\mathbf{F}_{\langle a,b ]}\psi$ are used to store the time elapsed since the occurrence of events involving $\psi$ between the current time instant $t$ and $t+b$.
In fact, \cite{MNP06} shows that formulae of the form $\mathbf{F}_{\langle a,b \rangle}\psi$ have inherent bounded variability (the result holds for signals with no l.c.r.o. restriction).

\begin{lemma}[\cite{MNP06}]\label{lemma-bounded-var-F}
Let $\theta=\mathbf{F}_{\langle a,b \rangle}\psi$, $\sig$ be a signal and let $0<t_1<t_2$ be two instants such that $\sig,t_1 \models \uparrow_\theta$, $\sig,t_2 \models \downarrow_\theta$ and $\forall t \in (t',t'') \ \sig,t \models \theta$.
Then, $t_2-t_1\geq b-a$.
\end{lemma}



By Lemma~\ref{lemma-bounded-var-F}, two consecutive events $\uparrow_\theta$ and $\downarrow_\theta$ for formulae $\theta=\mathbf{F}_{\langle a,b \rangle}\psi$ cannot occur at a distance less than $b-a$.
However, this does not hold when $\uparrow_\theta$ occurs at $t=0$ and $\psi$ is true at 0, but it becomes false before $b$. 
For instance, let $\sig,a \models p$ and $\sig,a+\varepsilon \models \downarrow_p$, where $\varepsilon>0$ is such that $a+\varepsilon< b$; assume for simplicity that $p$ remains false, i.e., for all $t\in [a+\varepsilon, +\infty)$, $\sig,t \not\models \psi$. 
Then, we have that $\sig,0 \models \uparrow_\theta$ and $\sig,\varepsilon \models \downarrow_\theta$.
This property will be exploited in Sect.~\ref{subsection-operators-sem} to define the translation of the $\mathbf{F}$ operator.

\begin{corollary}\label{cor-F-a-ge-0}
Let $\theta=\mathbf{F}_{\langle a,b ]}\phi$ be a MITL formula, with  $a>0$, $b\not= \infty$, and let $t$ be an instant of time. 
Then, in $[t,t+b]$ there are at most $d=2\left\lceil\frac{b}{b-a}\right\rceil$
 events $\uparrow_\theta,\downarrow_\theta$.
\end{corollary}
The result of Corollary~\ref{cor-F-a-ge-0} can be significantly simplified for formulae of the form $\theta=\mathbf{F}_{\langle 0,b ]}\phi$ or of the form $\theta=\mathbf{F}_{\langle a,+\infty )}\phi$.
In fact, in the former case, let $t_2>t_1\geq 0$ be two time instants such that $\sig, t_1\models \uparrow_\phi$, $\sig, t_2\models \downarrow_\phi$ and $\forall t'\in [t_2,t_2+b] \ \sig, t' \not\models \phi$.
Then, by definition, we have $\sig, t_1-b\models \uparrow_\theta$, $\sig, t_2\models \downarrow_\theta$ and $\forall t' \in [t_1-b,t_2) \ \models \theta$.
Therefore, no event for $\theta$ occurs over the interval $[t_1-b,t_2)$. 
If $\theta=\mathbf{F}_{\langle a,+\infty )}\phi$, by definition, $\sig,t \models \theta \iFF \exists t' \in \langle t+a, +\infty) \ \sig,t'\models \phi$; 
hence, $\sig,t \models \theta \Rightarrow \sig,0 \models \theta$, i.e., $\sig,0 \models \uparrow_\theta$. 
Event $\uparrow_\theta$ occurs in 0 if, and only if: $\exists t\geq a \ \sig,t \models \uparrow_\phi \text{or } \exists t> a \ \sig,t \models \downarrow_\phi \text{or } \exists t< a \ \sig,t \models \uparrow_\phi \wedge \forall t' > t \ \sig,t' \models \phi$.
Moreover, $\sig,t \not\models \theta \Rightarrow \forall t' \in \langle t+a, +\infty) \ \sig,t'\not\models \phi$, i.e., $\sig, t \models \downarrow_\theta \iFF \sig, t+a \models \downarrow_\phi \wedge \G{\neg \phi}$.
By the previous properties, the translation of formulae involving $\mathbf{F}_{\langle 0,b ]}$ and $\mathbf{F}_{\langle a,+\infty )}$ is simpler than the case $a>0$ and $b\not=\infty$, because auxiliary clocks are not needed to represent the formula.
For this reason, we provide a direct translation for these subformulae.

Since signals are finitely variable, all the events in $\sig$ 
can be enumerated as follows.
A position $i\geq 0$ uniquely identifies a time instant along $\sig$. 
Let $T\subset \Real_+$ be an infinite, but enumerable, set of time instants that includes 0 and every instant when at least one event occurs.
Let $I: T \to\Nat $ be a one-to-one mapping, 
consistent with the ordering of time, i.e, $I(0) = 0$ and $ I(t)<I(t')\iFF  t<t'$, and such that for all $t_1<t_2 \in T$ 
$I(t_2)=I(t_1)+1 \iFF \neg \exists t \, (t_1<t<t_2 \wedge t \in T)$.
By definition, for each subformula $\theta$ an event (either $\Hg{\theta}$ or $\Lw{\theta}$) always occurs at $I(0)=0$. 

Now, given a MITL formula $\phi$ and a signal $\sig$ such that $\sig, 0\models \phi$, we define how to build CLTL-oc interpretations 
from $\sig$.
We will prove afterwards that this interpretation
is a model for the CLTL-oc formula translating $\phi$.
We say that a clock $v$ is \textit{reset} at position $i$ when $\sigma(i,v)=0$.

Let $\pair{\pi}{\sigma}$ be a CLTL-oc interpretation. 
If an event for $\theta\in sub(\phi)$ occurs at $t\geq 0$, the corresponding event marker ($\Hg{\theta}$ or $\Lw{\theta}$) labels $\pi(I(t))$ and a reset for one of $z^0_\theta,z^1_\theta$ occurs at $I(t)$:
\begin{itemize}
\item $\bigvee_{i \in \{0,1\}} \sigma(I(t),z^i_\theta)=0$ and $\pair{\pi}{\sigma}, I(t) \models \Hg{\theta}$ if $\sig,t \models \uparrow_\theta$
\item $\bigvee_{i \in \{0,1\}} \sigma(I(t),z^i_\theta)=0$ and $\pair{\pi}{\sigma}, I(t) \models \Lw{\theta}$ if $\sig,t \models \downarrow_\theta$. 
\item $\sigma(0,z^0_\theta)=0$ for all $\theta$.
\item $\sigma(0,x_\theta^0)=0$ for all $\theta$ of the form $\mathbf{F}_{\langle a,b \rangle} \psi$.
\end{itemize}

Note that, by definition, for all time instants $t\in T$ where no events for $\theta$ occur, neither $\Hg{\theta}$ nor $\Lw{\theta}$ hold in $\pi(I(t))$ (i.e., $\pair{\pi}{\sigma}, I(t) \models \neg \Hg{\theta} \wedge \neg \Lw{\theta}$).

Now we define how CLTL-oc models represent time progress.
Let $t,t' \in T$ be two time instants such that $I(t')=I(t)+1$. 
For all clocks $z^i_\theta$ that are not reset in $I(t')$ we impose 
$$
\sigma(I(t'),z^i_\theta)=\sigma(I(t),z^i_\theta)+t'-t.
$$
In addition, $\exists i \in \{0,1\}$ s.t. $\sigma(I(t),z^i_\theta)= 0$ if and only if $\pair{\pi}{\sigma}, I(t) \models \Hg{\theta}$ or $\pair{\pi}{\sigma}, I(t) \models \Lw{\theta}$.
Clocks $z^0_\theta, z^1_\theta$ cannot be reset at the same time, but alternate, and $z^0_\theta$ is reset in the origin.
Clocks $x^j_\theta$ are dealt with analogously. As mentioned, there exist $d=2\left\lceil\frac{b}{b-a}\right\rceil$ clocks $x^j_\theta$ for a formula $\mathbf{F}_{\langle a,b \rangle}\psi \in sub(\phi)$.
First,  for all positions $i\geq 0$, $\sigma(i,z^0_\theta)=0$ or $\sigma(i,z^1_\theta)=0$ if, and only if, $\bigvee_{j=0}^{d-1} \sigma(i,x^j_\theta) = 0$, i.e, 
whenever an event for $\theta$ occurs, (at least) one auxiliary clock is reset. 
To avoid simultaneous resets of different clocks, if $x^j_\theta$ is reset then no $x^{j'}_\theta$ is reset, for $j'\not= j$. 
Auxiliary clocks are circularly reset  modulo $d$; i.e., if $x^j_\theta$ is reset at position $i$, then the next reset of $x^j_\theta$, if it exists, occurs in a position $i'>i$ such that all other clocks $x^{j'}_\theta$ ($j' \neq j$) are reset, in order, exactly once in $(i,i')$.
Note that, if a clock $x^j_\theta$ is reset at position $i = I(t)$, the next position $i' = I(t')$ when the clock is reset must be such that $t' > t+b$, i.e., given a formula $\theta=\mathbf{F}_{\langle a,b ]}$, every clock $x^j_\theta$ is reset only once over intervals of length $b$.
The sequence of resets starts with $x^0_\theta=0$.

Finally, if $\phi$ is satisfiable and $\sig$ is a signal such that $\sig,0 \models \phi$ i.e., $\sig,0 \models \uparrow_\phi$, then $\pair{\pi}{\sigma}, 0 \models \Hg{\phi}$.

Let $r_\phi(\sig)$ denote the (infinite) set of pairs $\pair{\pi}{\sigma}$ obtained from $\sig$ by means of the previous rules for a MITL formula $\phi$.
The inverse mapping $r_\phi^{-1}$ is also definable, but 
not all pairs $(\pi,\sigma)$ represent legal signals. Hence, we restrict them to the set of CLTL-oc models that are images of a signal $\sig$ under $r_\phi$, i.e., $(\pi,\sigma)$ is such that there exists a signal $\sig$ such that $(\pi,\sigma)\in r_\phi(\sig)$.
Sect. \ref{subsection-clocks-sem} provides a set of CLTL-oc formulae whose models are exactly the set of pairs $\pair{\pi}{\sigma}$ such that $\pair{\pi}{\sigma}\in r_\phi(\sig)$.
For these models the inverse map $r^{-1}$ is well-defined.

\subsection{Clocks and Events}\label{subsection-clocks-sem}
The following formulae define how events $\Hg{\theta}, \Lw{\theta}$ occur, for $\theta \in sub(\phi)$, and when clocks $z^0_\theta,z^1_\theta$ are reset.
However, they do not capture the semantics of subformulae $\theta$, which is the object of Sect. \ref{subsection-operators-sem}, but only the relations between events $\Hg{\theta}$ and $\Lw{\theta}$ and clock resets.

Formula \eqref{eq:clockreset} enforces that the occurrence of an event $\Hg{\theta}, \Lw{\theta}$ entails the reset of one of $z^0_\theta,z^1_\theta$.
In addition, Formula $z^0_\theta = 0$ evaluated in the origin states that clock $z^0_\theta$ is reset in the origin.
\begin{equation}
\Hg{\theta} \vee \Lw{\theta} \iFF z^0_\theta=0 \vee z^1_\theta = 0
\label{eq:clockreset}
\end{equation}

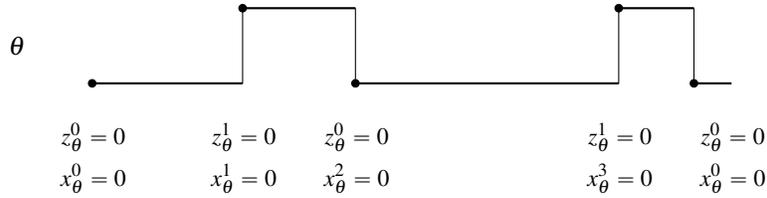
\begin{figure}\centering
\begin{tikzpicture}
\draw[thick]  (0,0) -- (2,0);
\draw[thick]  (2,1) -- (3.5,1);
\draw[thick]  (3.5,0) -- (7,0);
\draw[thick]  (7,1) -- (8,1);
\draw[thick]  (8,0) -- (8.5,0);

\draw  (2,0) -- (2,1);
\draw  (3.5,0) -- (3.5,1);
\draw  (7,0) -- (7,1);
\draw  (8,0) -- (8,1);

\draw[fill] (0,0) circle (0.05);
\draw[fill] (2,1) circle (0.05);
\draw[fill] (3.5,0) circle (0.05);
\draw[fill] (7,1) circle (0.05);
\draw[fill] (8,0) circle (0.05);

\draw node at (0,-1) (a) {\footnotesize$\begin{gathered}z_\theta^0=0 \\ x_\theta^0=0 \end{gathered}$};
\draw node at (2,-1) (b) {\footnotesize$\begin{gathered}z_\theta^1=0 \\ x_\theta^1=0 \end{gathered}$};
\draw node at (3.5,-1) (c) {\footnotesize$\begin{gathered}z_\theta^0=0 \\ x_\theta^2=0 \end{gathered}$};
\draw node at (7,-1) (d) {\footnotesize$\begin{gathered}z_\theta^1=0 \\ x_\theta^3=0 \end{gathered}$};
\draw node at (8.5,-1) (d) {\footnotesize$\begin{gathered}z_\theta^0=0 \\ x_\theta^0=0 \end{gathered}$};

\draw node at (-1,0.5) (a) {\small $\theta$};
\end{tikzpicture}
\caption{Sequence of circular resets for formula $\theta = \mathbf{F}_{\langle 2,1]} \psi$}\label{fig-circular}
\end{figure}

Let $a\in \Nat$ and value $\overline{a}_k$ be $(a\bmod k)$.
The clocks associated with a subformula $\theta$ are alternatively reset, as shown on an example in Figure~\ref{fig-circular}. Hence, 
between any two resets of clock $z^0_\theta$ there must be a reset of clock $z^1_\theta$, and vice-versa:
\begin{equation}
\label{eq:clockresetorder}
(\bigwedge_{i \in \{0,1\}} (z^i_\theta=0)) \Rightarrow \X{(z^{\overline{(i+1)}_2}_\theta=0) \R (z^i_\theta\neq 0)}.
\end{equation}
%

For a position $i>0$ it may happen that neither $\Hg{\theta}$ nor $\Lw{\theta}$ occur for any formula (i.e, no events occur).
The assumption that intervals are l.c.r.o. entails that intervals have non-null durations, and events $\uparrow_\theta,\downarrow_\theta$ cannot occur at the same time.
Define $\mathtt{events}_\phi  = \bigwedge_{\theta \in sub(\phi)} (z^0_\theta = 0) \wedge \Gltl{}{\eqref{eq:clockreset} \wedge \eqref{eq:clockresetorder}}$.

\begin{lemma}\label{lemma-clocks-event-r-inv}
Let $\theta$ be a symbol of a MITL formula.
For any non-Zeno signal $\sig: \Real_+ \to \{\emptyset, \theta\}$ for $\theta$ and for all $\pair{\pi}{\sigma} \in r_\theta(\sig)$, then $\pair{\pi}{\sigma}, 0 \models \mathtt{events}_\theta$.
Conversely, given $\pair{\pi}{\sigma}$ in which time is divergent and s.t. $\pair{\pi}{\sigma}, 0 \models \mathtt{events}_\theta$, there is exactly one non-Zeno signal $\sig$ s.t. $\sig = \rmapinv{\theta}{\pair{\pi}{\sigma}}$.
\end{lemma}



Let $\theta$ be $\mathbf{F}_{\langle a,b]} \psi$. 
We introduce $d=2\left\lceil\frac{b}{b-a}\right\rceil$ clocks $x^j_\theta$, which behave in a similar way as $z^0_\theta,z^1_\theta$.
Each $x^j_\theta$ is needed to store the time elapsed since the occurrence of the last event of $\theta$ ($\uparrow_\theta$ or $\downarrow_\theta$).
When one of $\uparrow_\theta, \downarrow_\theta$ occurs, then a $x^j_\theta$ is reset, i.e.,
$x^j_\theta=0$. 
Each reset event marked by $x^i_\theta=0$ entails either $\Hg{\theta}$ or $\Lw{\theta}$ and all $\uparrow_\theta, \;\downarrow_\theta$ events are marked by a single reset $x^i_\theta=0$ (Formula \eqref{clocks-event-reset}).
\begin{gather}
\label{clocks-event-reset}
\left(\Hg{\theta} \vee \Lw{\theta} \iFF \bigvee_{j=0}^{d-1}  x^j_\theta=0\right)
\ \wedge \
\left(\bigwedge_{i=0}^{d-1} \bigwedge_{j=0,i\not= j}^{d-1} \neg (x^i_\theta=0 \wedge x^j_\theta=0)\right)
\end{gather}
%
%
The occurrence of resets for clocks $x^i_\theta$ is circularly ordered and the sequence of resets starts from the origin by $x^0_\theta$ (see an example in Figure~\ref{fig-circular}).
If $x^i_\theta=0$, then, from the next position, all the other clocks are strictly greater than 0 until the next $x^{\overline{i+1}_d}_\theta=0$ occurs.

\begin{equation}\label{clocks-circularity}
\bigwedge_{i=0}^{d-1}\left(
x^i_\theta=0 \Rightarrow \X{
(x^{\overline{i+1}_d}_\theta=0)\R
\bigwedge_{j\in [0,d-1],\, j\not= i}(x^{\overline{j+1}_d}_\theta>0) } 
\right)
\end{equation}
Formula $x^0_\theta=0$, evaluated at position 0, sets the first reset of the sequence, constrained by formulae \eqref{clocks-event-reset}-\eqref{clocks-circularity}.
Moreover, we force all clock values to be strictly ordered in the origin by $x^0_\theta < x^{d-1}_\theta < \dots < x^1_\theta$, guaranteeing 
that resets are correctly associated with events occurring after the origin.

%
%

The following lemma (whose proof is similar to the one for Lemma \ref{lemma-clocks-event-r-inv}) shows that $\mathtt{auxclocks}_\theta$, defined as
$(x^0_\theta=0) \wedge \G{(\ref{clocks-event-reset}) \wedge (\ref{clocks-circularity}) }$ 
captures map $r$ for $\Fltl{\langle a,b]}{}$ formulae. .

\begin{lemma}\label{lemma-auxiliary-clocks}
Let $\theta = \mathbf{F}_{\langle a,b ]} \psi$.
For any signal $\sig: \Real_+ \to \{\emptyset, \theta\}$ for $\theta$ and for all $\pair{\pi}{\sigma} \in r_\theta(\sig)$,  it is $\pair{\pi}{\sigma}, 0 \models \mathtt{auxclocks}_\theta$.
Conversely, if $\pair{\pi}{\sigma}, 0 \models \mathtt{auxclocks}_\theta$, there exists one, and only one, signal $\sig$ s.t. $\sig = \rmapinv{\theta}{\pair{\pi}{\sigma}}$.
\end{lemma}


\subsection{Semantics of MITL Temporal Modalities}\label{subsection-operators-sem}
We now define a mapping $m$ associating a MITL formula with an equisatisfiable CLTL-oc formula, thus capturing the semantics of MITL in CLTL-oc.

The cases for Boolean connectives and the non-metric $\Ultl{}{}$ operator are straightforward.
In the following we write $\orig$ instead of $\neg\Y{\top}$ to represent the first position of CLTL-oc models.


%

\myitem{$\theta=p \in AP$} it follows from the definition of $\Hg{p}$ and $\Lw{p}$, representing events $\uparrow_p, \downarrow_p$ over discrete time.
%

\myitem{$\theta= \neg\psi$} in this case it is
$
m(\theta)=
\allup{\theta} \iFF \alldn{\psi}
$.


\myitem{$\theta=\gamma\wedge\psi$} we have:
$
m(\theta) = \allup{\theta}\iFF \allup{\gamma} \wedge \allup{\psi}
$.

\myitem{$\theta=\gamma\U_{(0,+\infty)}\psi$} similarly:
$
m(\theta) = \allup{\theta} \iFF \allup{\gamma} \wedge \allup{\gamma} \U \allup{\psi}
$.

\myitem{$\theta=\mathbf{F}_{\langle a, b]} \psi$}
When an event $\uparrow_\theta$ occurs, a clock $x^j_\theta$ is
reset, then event $\uparrow_\psi$ will 
eventually occur after $b$ time units
and it has to occur after $b-a$ instants from the last occurrence of $\downarrow_\psi$ (otherwise $\uparrow_\theta$ has already occurred in the past).
The case for $t=0$ is treated separately: $\uparrow_\theta$ occurs at 0 when there is an interval in which $\psi$ holds that either starts in $[a,b]$ or it spans $a$.
Clock $x^0_\theta$ is used to measure the time elapsing from the origin.
In fact, by Corollary \ref{cor-F-a-ge-0}, $x^0_\theta$, which is reset at 0, can only be reset again after $b$.
%
\begin{equation}
\Hg{\theta} \iFF
\begin{gathered}
\neg \orig \wedge
\bigvee_{j=0}^{d-1}(x^j_\theta=0) \wedge \Xltl{}\left({x^j_\theta>0 \, \U \left(
                                                                       \Hg{\psi} \ \wedge x^j_\theta=b
                                                                       \wedge \bigvee_{i \in \{0,1\}} z^i_\psi>(b-a)
                                                                       \!\right)}\!\!\right)
\ \ \vee \\
\orig \wedge 
{(\orig \vee x^0_\theta>0) \, \U
\left(
\begin{gathered}
\allup{\psi} \wedge
\left(
  a \leq x^0_\theta \leq b \ \vee
  x^0_\theta < a \wedge \X{x^0_\theta > a}
\!\right)
\end{gathered}\!\!\right)} 
\end{gathered}\label{sem-Fup}
\end{equation}

\begin{figure}[h!]\centering
\begin{tikzpicture}
\draw[thick]  (2.2,0.5) -- (3,0.5);
\draw[thick, dashed] (1.4,0.5) -- (2.2,0.5);
\draw[thick]  (0.6,0.5) -- (1.4,0.5);
\draw[thick]  (3,1) -- (4,1);
\draw[thick, dashed] (4,1) -- (4.6,1);

\draw[thick]  (-0.5,-2) -- (0,-2);
\draw[thick]  (0,-1.5) -- (0.5,-1.5);
\draw  (0.6,0.5) -- (0.6,1);
\draw[thick, dashed] (0.6,-1.5) -- (1,-1.5);
\draw[thick, dashed] (-1,-2) -- (-0.6,-2);

\draw (3,0.5) -- (3,1);

\pgfsetarrowsstart{|};
\draw[dashed] (0,-0.5) -- (1.8,-0.5);
\draw node at (3,-0.1) (a) {\small $b$};
\draw node at (1.85,-0.1) (a) {\small $a$};

\pgfsetarrowsend{]};
\pgfsetarrowsstart{to};
\draw[very thick, dashed] (1.8,-0.5) -- (3,-0.5);

\draw[-] (0,-1.5) -- (0,-2);

\draw[fill] (0.6,0.5) circle (0.05);
\draw[fill] (0,-1.5) circle (0.05);
\draw[fill] (3,1) circle (0.05);

\draw node at (-1.6,0.5) (a) {\small $\psi$};
\draw node at (-2.3,-2) (a) {\small $\theta=\mathbf{F}_{\langle a, b]} \psi$};

\draw node at (0,-1) (a) {\footnotesize $x^i_\theta=0$};
\draw node at (3,-1) (a) {\footnotesize $x^i_\theta=b$};

\draw node at (0.5,1.5) (a) {\footnotesize $z^i_\theta=0$};
\draw node at (3,1.5) (a) {\footnotesize $z^i_\theta>b-a$};


\end{tikzpicture}
\caption{Rising edge }\label{fig-rising}
\end{figure}
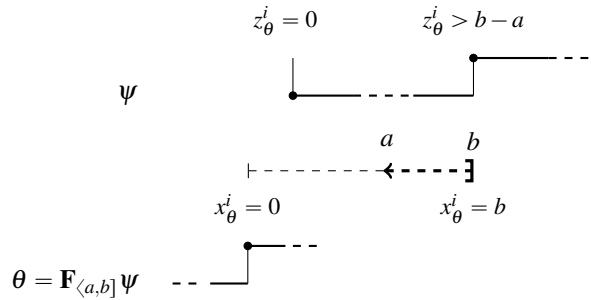

Formula \eqref{sem-Fupconstr} defines the condition to make $\Hg{\theta}$ true exactly $b$ instants before an event $\Hg{\psi}$, provided that clock $z^i_\psi$ 
is greater than 
$(b-a)$ when $\Hg{\psi}$ occurs (i.e., the last time $\psi$ became false was at least $b-a$ time units before). 
An illustration of Formulae~\eqref{sem-Fup} and \eqref{sem-Fupconstr} is in Figure~\ref{fig-rising}.
\begin{equation}
\label{sem-Fupconstr}
\Hg{\psi} \wedge \bigvee_{i \in \{0,1\}} z^i_\psi>(b-a) \Rightarrow\bigvee_{j=0}^{d-1} x^j_\theta=b
\end{equation}

When an event $\downarrow_\theta$ occurs, a clock $x^j_\theta$ is
reset, then the event $\downarrow_\psi$ will eventually occur after exactly $a$ time units
and the next $\uparrow_\psi$ cannot occur before another $b-a$ instants after that (otherwise $\downarrow_\theta$ cannot occur).
In the origin, however, $\downarrow_\theta$ occurs also in the case that $\uparrow_\theta$ does not occur.
\begin{equation}\label{sem-Fdown}
\Lw{\theta} \iFF 
\begin{gathered}
\bigvee_{j=0}^{d-1}
  (x^j_\theta=0) \wedge
  \X{ (x^j_\theta>0) \U
  \left(
  \Lw{\psi} \wedge x^j_\theta=a \ \wedge
  \Hg{\psi}\R \neg \left(\Hg{\psi} \wedge x^j_\theta \leq b \right)
  \right)}
\ \ \vee \ \ 
(\orig \wedge \neg \Hg{\theta})
\end{gathered}
\end{equation}

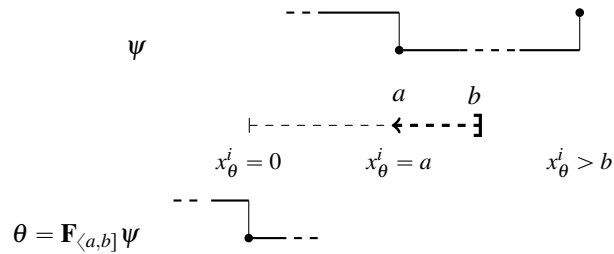
\begin{figure}[h!]\centering
\begin{tikzpicture}
\draw[thick,dashed]  (2.5,1) -- (3,1);
\draw[thick, dashed] (4.8,0.5) -- (5.6,0.5);
\draw[thick]  (5.6,0.5) -- (6.4,0.5);
\draw[thick]  (4,0.5) -- (4.8,0.5);
\draw[thick]  (3,1) -- (4,1);

\draw[thick]  (1.5,-1.5) -- (2,-1.5);
\draw[thick,dashed]  (1,-1.5) -- (1.4,-1.5);
\draw[thick]  (2,-2) -- (2.5,-2);
\draw[thick,dashed]  (2.6,-2) -- (3,-2);

\draw  (4,0.5) -- (4,1);
\draw  (6.4,0.5) -- (6.4,1);

\pgfsetarrowsstart{|};
\draw[dashed] (2,-0.5) -- (3.8,-0.5);
\draw node at (5,-0.1) (a) {\small $b$};
\draw node at (4,-0.1) (a) {\small $a$};
\pgfsetarrowsend{]};
\pgfsetarrowsstart{to};
\draw[very thick, dashed] (3.9,-0.5) -- (5.1,-0.5);

\draw[-]  (2,-1.5) -- (2,-2);

\draw[fill] (6.4,1) circle (0.05);
\draw[fill] (2,-2) circle (0.05);
\draw[fill] (4,0.5) circle (0.05);

\draw node at (0.5,0.5) (a) {\small $\psi$};
\draw node at (-0.3,-2) (a) {\small $\theta=\mathbf{F}_{\langle a, b]} \psi$};

\draw node at (2,-1) (a) {\footnotesize $x^i_\theta=0$};
\draw node at (4,-1) (a) {\footnotesize $x^i_\theta=a$};
\draw node at (6.4,-1) (a) {\footnotesize $x^i_\theta>b$};


\end{tikzpicture}
\caption{Falling edge }\label{fig-falling}
\end{figure}

Formula \eqref{sem-Fdownconstr} is the dual of \eqref{sem-Fupconstr} for a falling edge (Figure~\ref{fig-falling}); 
it defines a sufficient condition forcing $\Lw{\theta}$ when an event $\Lw{\psi}$ occurs and $\Hg{\psi}$ does not happen before $(b-a)$ time units have passed since $\Lw{\psi}$.
\begin{equation}\label{sem-Fdownconstr}
\Lw{\psi}\wedge \Hg{\psi} \R \neg \left( \Hg{\psi} \wedge \bigwedge_{i \in \{0,1\}} z^i_\psi\leq (b-a) \right) \Rightarrow\bigvee_{j=0}^{d-1} x^j_\theta=a
\end{equation}

Formula $m(\theta)$ in this case is $\eqref{sem-Fup} \wedge \eqref{sem-Fupconstr} \wedge \eqref{sem-Fdown} \wedge \eqref{sem-Fdownconstr}$.

\medskip
As already anticipated, we may study separately the case of formulae $\mathbf{F}_{\langle a, b]} \psi$ where $a=0$ or $b = +\infty$.
The translation in the two cases is simpler than the general one
because auxiliary clocks are no longer required to measure the time elapsing between events involving signal for the formula.

\myitem{$\theta=\mathbf{F}_{\langle 0, b]} \psi$} the translation for event $\uparrow_\theta$ is analogous to the one of the general case where time elapsing is measured with respect to the clock $z^j_\theta$ that is reset when $\Hg{\theta}$ holds (recall that, by Corollary \ref{cor-F-a-ge-0}, $z^j_\theta$ can be reset again only after the occurrence of $\Hg{\psi}$).
The semantics of $\downarrow_\theta$ in this case is simpler than for Formula (\ref{sem-Fdown}) because
events $\downarrow_\psi$ and $\downarrow_\theta$ always occur simultaneously, provided that the next $\uparrow_\psi$ does not occur within $b$ time instants from $\downarrow_\psi$.
\begin{equation}
\label{sem-Fup0b}
\Hg{\theta} \iFF
\left(
\begin{gathered}
\neg \orig \wedge \alldn{\psi} \wedge
\left(\bigvee_{j = 0}^{1}(z^j_\theta=0) \wedge \Xltl{}\left({z^j_\theta>0 \, \U \left(
                                                                       \Hg{\psi} \ \wedge z^j_\theta=b
                                                                       \wedge \bigvee_{i \in \{0,1\}} z^i_\psi>b
                                                                       \!\right)}\!\!\right)\right)
\ \ \vee \\
\orig \wedge 
{(\orig \vee z^0_\theta>0) \, \U
(
\allup{\psi} \wedge  z^0_\theta \leq b
)} 
\end{gathered}
\right)
\end{equation}
\begin{equation}
\label{sem-Fupconstr0b}
\Hg{\psi} \wedge \bigvee_{i \in \{0,1\}} z^i_\psi>b \Rightarrow\bigvee_{j \in \{0,1\}} z^j_\theta=b
\end{equation}
\begin{equation}\label{sem-Fdown0b}
\Lw{\theta} \iFF
\Lw{\psi}
\wedge 
  \Hg{\psi}\R \neg \left( \Hg{\psi} \wedge \bigwedge_{i \in \{0,1\}} z^i_\psi\leq b \right)
\end{equation}

\myitem{$\theta=\mathbf{F}_{\langle a, +\infty)} \psi$} From the semantics of $\Fltl{a,+\infty} \psi$ it is easy to see that event $\uparrow_\theta$ may only occur at 0, if $\psi$ eventually holds in the future after $a$ instants from the origin.
Similarly, event $\downarrow_\theta$ may only occur once, but not necessarily in the origin; 
more precisely, it holds at 0 if and only if $\uparrow_\theta$ does not hold at 0, while for every instant $t>0$ it occurs when event $\downarrow_\psi$ occurs in $t+a$ and $\psi$ is always false afterwards.
As a consequence, $z^1_\theta$ is reset at most once, if $\Lw{\theta}$ occurs in an instant other than the origin.
\begin{equation}
\label{sem-Fupainf}
\Hg{\theta} \iFF
\orig \wedge 
{(\orig \vee z^0_\theta>0) \, \U
\left(\!
\begin{gathered}
\allup{\psi} \wedge
\left(\!
  a \leq z^0_\theta \ \vee
  z^0_\theta < a \wedge \X{z^0_\theta > a}
\!\right)
\end{gathered}\!\!\right)} 
\end{equation}
\begin{equation}\label{sem-Fdownainf}
\Lw{\theta} \iFF
  z^1_\theta=0 \wedge
  \Xltl{}\left({ z^1_\theta>0 \ \U
  \left(
  \Lw{\psi} \wedge z^1_\theta=a \ \wedge
  \Gltl{}{\neg \Hg{\psi}}
  \right)}\!\!\right)
\ \ \vee \ \
(\orig \wedge \neg \Hg{\theta})
\end{equation}
\begin{equation}\label{sem-Fdownconstrainf}
\Lw{\psi}\wedge \Gltl{}{\neg \Hg{\psi}} \Rightarrow z^1_\theta=a
\end{equation}


\subsection{Correctness}
\label{subsec:correctness}

Let $F$ be a set of formulae.
We extend map $r$ to $sub(\phi)$, written $\rmap{sub(\phi)}{\sig}$, to represent the set of CLTL-oc models where atomic propositions are  symbols associated with each subformula in $\phi$ and variables includes all clocks $z^0_\theta,z^1_\theta$ and the auxiliary clocks for the case  $\mathbf{F}_{\langle a,b ]}$.
\begin{lemma}\label{lemma-map-formulae1}
Let $\sig$ be a signal, and $\phi$ a MITL formula.
For any $\pair{\pi}{\sigma} \in \rmap{sub(\phi)}{\sig}$ it is:
\[
\pair{\pi}{\sigma}, 0 \models \bigwedge_{\theta\in sub(\phi)} \G{m(\theta)} \;  \wedge \mathtt{events}_\theta\wedge \bigwedge_{\stackrel{\theta\in sub(\phi)}{\theta=\mathbf{F}_{\langle a,b ]}}}\mathtt{auxclocks}_\theta
\]
and for all $k \in \Nat, \theta\in sub(\phi)$ it is $\pair{\pi}{\sigma}, k \models m(\theta)$.
\end{lemma}

\begin{lemma}\label{lemma-map-formulae2}
Let $\sig$ be a signal and let $\phi$ be a MITL formula.
If
$$
\pair{\pi}{\sigma}, 0 \models \bigwedge_{\theta\in sub(\phi)} \G{m(\theta)}  \wedge \mathtt{events}_\theta\wedge \bigwedge_{\stackrel{\theta\in sub(\phi)}{\theta=\mathbf{F}_{\langle a,b ]}}}\mathtt{auxclocks}_\theta
$$
and $\sig = \rmapinv{sub(\phi)}{(\pi, \sigma)}$, then for all $t \in T$ it is
$\pair{\pi}{\sigma}, I(t) \models \Hg{\phi}$ iff $\sig, t \models \uparrow_{\phi}$ (similarly for $\Lw{\phi}$).
\end{lemma}

The main result, the equisatisfiability of MITL and of its CLTL-oc translation, follows.
\begin{theorem}\label{theorem-equisat}
A MITL formula 
$\phi$ is satisfiable if, and only if the following formula is satisfiable:
\begin{equation}
\label{eq:transform}
\Hg{\phi}
\wedge \bigwedge_{\theta \in sub(\phi)} \G{m(\theta)} \wedge \mathtt{events}_\phi \wedge \bigwedge_{\stackrel{\theta\in sub(\phi)}{\theta=\mathbf{F}_{\langle a,b ]}}}\mathtt{auxclocks}_\theta.
\end{equation}
\end{theorem}

\subsection{Complexity}
\label{subsec:complexity}

The reduction of MITL  to CLTL-oc  of Sect.~\ref{subsection-operators-sem} induces an EXPSPACE decision procedure for the satisfiability of MITL (the problem is actually EXPSPACE-complete).
In fact, consider a MITL formula $\varphi$, and its CLTL-oc translation \eqref{eq:transform} obtained following the reduction of Sect.~\ref{subsection-operators-sem}.
In Formula \eqref{eq:transform} we introduce two clocks for each subformula of $\varphi$, unless the subformula is of the form $\mathbf{F}_{\langle a, b]} \psi$, in which case we introduce at most $b$ clocks, since $a,b \in \Nat$.
Then, the size of \eqref{eq:transform} is $O(|\varphi|K)$, where $K$ is the maximum constant appearing in $\varphi$.
It can be shown that satisfiability for a CLTL-oc formula $\phi_{\text{CLTL}}$ is PSPACE in the number of subformulae of $\phi_{\text{CLTL}}$ (which is $O(|\varphi|K)$ for Formula $\eqref{eq:transform}$) and in the size of the string encoding the maximum constant occurring in it ($K$ for Formula $\eqref{eq:transform}$).
Hence, the decision procedure induced by our encoding is in EXPSPACE when using a binary encoding of $K$.
As remarked in~\cite{AFH96}, if the MITL formula $\varphi$ does not contain subformulae of type $\mathbf{F}_{\langle a, b]} \psi$ (with $a > 0$ and $b \not = \infty$), the reduction of Sect.~\ref{subsection-operators-sem} only introduces one clock variable for each subformula.
As a consequence, the size of Formula \eqref{eq:transform} is $O(|\varphi|)$ and the algorithm is in PSPACE.

%% file: section-generalization.tex
\section{Generalized translation}\label{section-generalization}
Our translation from MITL to CLTL-oc can be extended to represent general signals where no assumption is made on their shape, other than their finite variability,
i.e., the l.c.r.o. assumption of Sect. \ref{section-reduction} can be relaxed. 
In this more general case, the truth of a formula $\phi$ can change in a \emph{singular} manner, that is, there can be instants where the value of $\phi$ is different than in a neighborhood thereof.

More precisely, we say that in a time instant $t$ of a signal $\sig$ formula $\phi$ has an ``up-singularity'' $\qtlSup{\phi}$ if it holds in $t$, but not before and after it; more precisely, we say that $\sig, t \models \qtlSup{\phi}$ if and only if $\sig, t \models \phi$ and $\exists \varepsilon>0$ s.t. $\forall t'\neq t \in (t-\varepsilon,t+\varepsilon)$ it is $\sig, t' \not\models \phi$. 
We say that $\phi$ has a ``down-singularity'' $\qtlSdn{\phi}$ when $\neg \phi$ has an up-singularity (i.e., $\phi$ does not hold in $t$, but it does before and after it).
Note that, by their definition, singularities (either up or down), cannot occur in $t=0$.

To represent general signals in CLTL-oc we ``split'' the representation of the value of subformulae $\theta$ in intervals $[t,t')$ in two parts: $\fst{\theta}$ captures the value of $\theta$ in $t$, whereas $\rest{\theta}$ corresponds to its value in $(t,t')$.
With the new predicates, we can restrict represented signals to only include l.c.r.o. intervals by imposing the constraint $\quad \fst{\theta} \iFF \rest{\theta} \quad$ for all $\theta$.
%
In addition, $\allup{\theta}$ and $\alldn{\theta}$ become: 
$
\hspace{0.5cm}
\allup{\theta} = \fst{\theta} \wedge \rest{\theta}
\hspace{1cm}
\alldn{\theta} = \neg \fst{\theta} \wedge \neg \rest{\theta}
$.
Then, the encoding of Sect.~\ref{section-reduction} can be used also with the new atomic predicates, provided constraint $\quad \fst{\theta} \iFF \rest{\theta} \quad$ is added for all subformulae.
If, instead, general signals are to be allowed, the encoding must be extended to include also the cases in which the values of (sub)formulae change in singular manners.

To this end, we slightly modify the definition of $\Hg{\xi}$ as $\neg \Yltl{}(\rest{\xi})\wedge \rest{\xi}$ and $\Lw{\xi}$ as $\neg \Yltl{}(\neg\rest{\xi})\wedge\neg\rest{\xi}$ and we introduce the following abbreviations, which capture, respectively, up- and down-singularities (note that neither $\Sup{\xi}$, nor $\Sdn{\xi}$ hold at 0, as $\Yltl{\bullet}$ is false there):
$$
\Sup{\xi} = \Yltl{}({\neg \rest{\xi}}) \wedge \fst{\xi} \wedge \neg \rest{\xi}
\hspace{1.1cm}
\Sdn{\xi} = \Yltl{}({\rest{\xi}}) \wedge \neg \fst{\xi} \wedge \rest{\xi}
$$
We also define the following:
$
\hspace{0.8cm}
\befDnowU{\xi}{} = \Hg{\xi} \vee \Sup{\xi} \vee (O\wedge \fst{\xi})
\hspace{1cm}
\nowonD{\xi}{} = \Lw{\xi} \vee \Sup{\xi} 
$.
\\
More precisely, $\befDnowU{\xi}{}$ corresponds to a situation where $\xi$ does not hold the interval before the current one (if such interval exists), and it is true sometimes in the current one (either in its first instant, in which case $\xi$ can have a up-singularity, or in the rest of the interval).
Dually, $\nowonD{\xi}{}$ holds if $\xi$ is true in the first instant of the current interval, or in the interval before it, and from that moment on it is false.

When general signals are allowed, there is no need to restrict the temporal operators only to  $\Fltl{\langle a,b]}{\psi}$.
For simplicity, we focus on the encoding of case $\theta = \Fltl{(a,b)}{\psi}$, all other cases being similar.

\myitem{$\theta=\Fltl{(a,b)}{}{\psi}$}
We have the following result.

\begin{lemma}
\label{lemma:FUpRLCont}
If $\theta = \Fltl{(a,b)}{}{\psi}$ is a MITL formula and $\sig, t \models \theta$ then s $\exists \varepsilon \in \Real_{>0}$ such that,
for all $t' \in [t,t+\varepsilon]$ it is $\sig, t' \models \theta$
and, when $t > 0$, there is also $\varepsilon \in \Real_{>0}$ such that $\varepsilon < t$ and for all $t' \in [t-\varepsilon, t]$ it is $\sig, t' \models \theta$.
\end{lemma}
%

Because of Lemma \ref{lemma:FUpRLCont}, an up-singularity $\Sup{\theta}$ can never occur for $\theta = \Fltl{(a,b)}{}{\psi}$.
In addition, if $\theta$ holds at the beginning of an interval (i.e., $\fst{\theta}$ holds), then it must hold also in the rest of the interval and, if $t > 0$, it must also hold in the interval before.
%
Then, the following constraint holds in every instant:
\begin{equation}
\label{sem-Fconstrgen}
\fst{\theta} \Rightarrow \rest{\theta} \wedge (\Yltl{}(\rest{\theta}) \vee \orig)
\end{equation}
Formula \eqref{sem-Fupgen} is similar to \eqref{sem-Fup}, but it specifies that, when $\theta$ becomes true outside of the origin, it must do so in a left-open manner (i.e., $\fst{\theta}$ does not hold with $\Hg{\theta}$); also, there is one additional condition that makes $\theta$ become true in 0 when $\psi$ becomes true exactly at $b$, in which case $\theta$ does not hold in 0.
\begin{equation}
\label{sem-Fupgen}
\Hg{\theta} \iFF
\begin{gathered}
\neg \orig \wedge
\neg \fst{\theta} \wedge
\bigvee_{j=0}^{d-1}(x^j_\theta=0) \wedge \Xltl{}\left({x^j_\theta>0 \, \U \left(
                                                                       \befDnowU{\psi}{} \ \wedge x^j_\theta=b
                                                                       \wedge \bigvee_{i=0}^1 z^i_\psi>(b-a)
                                                                       \!\right)}\!\!\right)
\ \ \vee \\
\orig \wedge
\neg \fst{\theta} \wedge
\Xltl{}\left({x^0_\theta>0 \, \U \left( \befDnowU{\psi}{} \ \wedge x^0_\theta=b
                                        \wedge \bigvee_{i=0}^1 z^i_\psi\geq (b-a)
                                        \!\right)}\!\!\right)
\ \ \vee \\
\orig \wedge \fst{\theta} \wedge 
{(\orig \vee x^0_\theta>0) \, \U
\left(
\begin{gathered}
  (\fst{\psi} \vee \rest{\psi}) \wedge a < x^0_\theta < b
  \quad \vee \quad
  \rest{\psi} \wedge x^0_\theta < a \wedge \X{x^0_\theta > a}
\end{gathered} \right)} 
\end{gathered}
\end{equation}
%
Formulae \eqref{sem-Fupconstrgen}, \eqref{sem-Fdowngen} and \eqref{sem-Fdownconstrgen} generalize, respectively, \eqref{sem-Fupconstr}, \eqref{sem-Fdown} and \eqref{sem-Fdownconstr} to include also the case in which $\psi$ changes its value in a singular manner (i.e., with $\Sup{\psi}$ instead of $\Hg{\psi}$ or $\Lw{\psi}$).
\begin{equation}
\label{sem-Fupconstrgen}
\befDnowU{\psi}{} \wedge \bigvee_{i \in \{0,1\}} z^i_\psi \geq (b-a) \Rightarrow\bigvee_{j=0}^{d-1} x^j_\theta=b
\end{equation}
\begin{equation}\label{sem-Fdowngen}
\Lw{\theta} \iFF
\bigvee_{j=0}^{d-1}
  (x^j_\theta=0) \wedge
  \Xltl{}\!\!\left(\!\!{ (x^j_\theta>0) \U\!
  \left(\!\!
  \begin{gathered}
  \nowonD{\psi}{} \wedge x^j_\theta=a \ \wedge
  \Xltl{}\!\!\left(\befDnowU{\psi}{}\R \neg \!\!\left(\befDnowU{\psi}{} \wedge x^j_\theta \leq b\right)\!\!\right)
  \end{gathered}\!\!\right)}\!\!\right)
\ \ \vee \ \
(\orig \wedge \neg \Hg{\theta})
\end{equation}
\begin{equation}\label{sem-Fdownconstrgen}
\nowonD{\psi}{} \wedge \Xltl{\befDnowU{\psi}{} \R \neg \left(\befDnowU{\psi}{} \wedge \bigwedge_{i=0}^1 z^i_\psi\leq (b-a) \right)} \Rightarrow\bigvee_{j=0}^{d-1} x^j_\theta=a
\end{equation}

Finally, we need to consider an additional shape in which $\theta$ can change value.
More precisely, there is also the case in which $\theta$ becomes false with a down-singularity $\Sdn{\theta}$.
This occurs in an instant $t$ (which must be $>0$, as singularities cannot occur in the origin by definition) such that $\psi$ becomes false at $t+a$, but it becomes true again at $t+b$ (and it stays false in interval $(t+a,t+b)$).
This condition is captured by Formula \eqref{sem-Fsinggen}, which is similar to Formula \eqref{sem-Fdowngen}, except that it specifies that when $\psi$ becomes true again, the clock $x^j_\theta$ that is reset when $\phi$ has the singularity has value $b$.
\begin{equation}\label{sem-Fsinggen}
\Sdn{\theta} \iFF \neg \orig \wedge 
\bigvee_{j=0}^{d-1}
  (x^j_\theta=0) \wedge
  \Xltl{}\!\!\left(\!\!{ (x^j_\theta>0) \U\!
  \left(\!\!
  \nowonD{\psi}{} \wedge x^j_\theta=a \ \wedge
  \Xltl{}\!\!\left(\notbefDnowU{\psi}{}\U \!\!\left(\befDnowU{\psi}{} \wedge x^j_\theta = b\right)\!\!\right)
  \!\!\right)}\!\!\right)
\end{equation}

Then, 
$m(\theta)$ is $\eqref{sem-Fconstrgen} \wedge \eqref{sem-Fupgen} \wedge \eqref{sem-Fupconstrgen} \wedge \eqref{sem-Fdowngen} \wedge \eqref{sem-Fdownconstrgen} \wedge \eqref{sem-Fsinggen}$.

\medskip

To allow for signals of general shape, the encoding for subformulae of the form $\Uinfix{(0,+\infty)}{\gamma}{\psi}$ must also be revisited.
As this is rather straightforward, we skip the details for reasons of brevity.
Instead, we point out that it is possible to define a CLTL-oc encoding also for MITL \emph{past} operators $\Sltl{}{}$ and $\Pltl{\langle a,b \rangle}{}$.
It is known that past operators increase the expressiveness of MITL \cite{BCM10}, but do not impact on decidability.
Hence, a decision procedure that also includes the possibility to handle past operators is more powerful than one dealing with the future-only fragment.
To conclude this section, we show the encoding $m(\theta)$ for the $\Sltl{}{}$ operator (whose semantics is symmetric to the one of $\Ultl{}{}$ shown in Table \ref{tab:MITLsemantics}).
The case for operator $\Pltl{\langle a,b \rangle}{}$ is omitted for brevity.

%
%

\myitem{$\theta=\Sinfix{(0,+\infty)}{\gamma}{\psi}$}
In this case it can be shown that, if $\sig$ is a finitely variable signal and $\theta$ holds in an instant $t$, then it must also hold in $(t-\varepsilon, t)$, for some $\varepsilon > 0$, and vice-versa.
Then, in $t=0$ $\theta$ is false, and there $\Snc$ formulae cannot have singularity points.
In addition, when a $\Snc$ formula changes its value after the origin, it must do so in a left-open manner (i.e., the value at the changing point is the same as the one before the changing point).
Then, we have 
\begin{equation}
\label{eq:semSince}
m(\theta) =
( \fst{\theta} \iFF \Yltl{}({\rest{\theta}}) )
\wedge
(
  \rest{\theta}
  \iFF 
      \allup{\gamma}\Sltl{}{}((\fst{\psi} \vee \rest{\psi}) \wedge \rest{\gamma})
).
\end{equation}

%% file: section-conclusions.tex
\section{Conclusions}
\label{section-conclusions}
This paper investigates a bounded approach to satisfiability checking of the continuous-time temporal logic MITL. 
We showed an encoding of MITL into a decidable logic (CLTL-oc), 
which allows, both in principle and in practice, 
the use of SMT solvers to check satisfiability of MITL.

A decision procedure  for CLTL-oc \cite{BRS13b} is 
implemented in a plugin, called  $\mathtt{ae^2zot}$, of our Zot toolkit~\cite{ae2zot},
whereas the reduction outlined in Sect.~\ref{section-reduction} and \ref{section-generalization} is implemented in the $\mathtt{qtlsolver}$ tool, available from~\cite{qtlsolver}.
The tool translates MITL (or the expressively equivalent QTL logic \cite{HR04})  into CLTL-oc, which can be checked for satisfiability by $\mathtt{ae^2zot}$.
The resulting toolkit has a 3-layered structure, where CLTL-oc is the intermediate layer between SMT-solvers and various temporal formalisms that can be reduced to CLTL-oc.
This not only supports (bounded) satisfiability verification of different languages, but it also allows the expression of different degrees of abstraction.
For instance, MITL abstracts away the notion of clocks, inherently encompassed within temporal modalities, 
which are instead explicit  in CLTL-oc and actually available to a user, e.g., 
to express or verify properties where clocks are convenient.
In fact, preliminary experimental results point out that the time required to solve CLTL-oc may be significantly 
smaller than the one needed for more abstract languages, such as MITL. This is caused by the ``effort'' required to capture the semantics of temporal modalities,
which, on the other hand, allow for more concise and manageable high-level specifications. 
This layered structure also allows the resolution of a formula to be compliant with constraints imposed at lower layers, for instance by adding at the   CLTL-oc layer some extra formula limiting the set of valid models (e.g., 
by discarding certain edges of some events or by adding particular timing requirements). 
Also the third layer (the SMT solver) may be used to add further constraints,
e.g., to force the occurrence of a proposition or of a certain clock value at a specific discrete position of the finite model.

The current implementation of $\mathtt{qtlsolver}$ supports the MITL-to-CLTL-oc translation, 
both with or without the l.c.r.o. restriction. In fact, 
the following encodings are currently available:
\begin{itemize}
\item[MITL] providing a direct definition of MITL operators, assuming l.c.r.o. intervals;

\item[QTL] providing the definition of generalized QTL operators (e.g., $\Fltl{(0,b)}{}$, $\Pltl{(0,b)}{}$) with unrestricted signals (other than they be finitely variable), and MITL operators through abbreviations.
\end{itemize}

We used the above two encodings to carry out some experiments (available from the $\mathtt{qtlsolver}$ website \cite{qtlsolver}, or  described in \cite{BRS13b}).
Let us illustrate one of them.
MITL Formula \eqref{eq:Spikes} specifies that predicate $p$ occurs in isolated points with a period of $100$ (i.e., it occurs exactly at 0, 100, 200, etc.).
\begin{equation}
\label{eq:Spikes}
\Gltl{[0,\infty)}{}
\left(
\left( \Gltl{(0,100)}{\neg p} \Rightarrow \Gltl{(100,200)}{\neg p} \right)
\wedge
(p \Rightarrow \Fltl{(0,200)}{p})
\right)
\wedge
p \wedge \Gltl{(0,100)}{\neg p}
\end{equation}
$\mathtt{qtlsolver}$ was able to find a model for Formula \eqref{eq:Spikes} in around 10 seconds, using a bound of 10.\footnote{All tests have been carried out on a desktop computer with a 2.8GHz AMD PhenomTMII processor and 8MB RAM; the solver was Microsoft Z3 3.2. The encoding used was the one for QTL, with unrestricted signals.}
Note that, even if the constants appearing in Formula \eqref{eq:Spikes} are in the order of the hundreds, events in the corresponding models occur only sparsely, hence a bound of 10 is enough for $\mathtt{qtlsolver}$ to satisfy \eqref{eq:Spikes}.
If we add to the specification Formula \eqref{eq:SpikesPast}, which states that $q$ must hold within 1 time unit in the past or in the future of each $p$, the solver finds a model (again, with bound 10) in about 40 seconds.
\begin{equation}
\label{eq:SpikesPast}
\Gltl{(0,\infty)}{p \Rightarrow \Fltl{(0,1)}{q} \vee \Pltl{(0,1)}{q} }
\end{equation}
Formula \eqref{eq:SpikesPast} does not impose that $q$ be false in between occurrences of $p$.
A more restricted behavior is obtained by adding also constraint \eqref{eq:SpikesPastqPer}, which imposes that $q$ occurs only in isolated instants, and that there must be at least $100$ time units between consecutive occurrences of $q$.
\begin{equation}
\label{eq:SpikesPastqPer}
\Gltl{(0,\infty)}{q \Rightarrow \Gltl{(0,100)}{ \neg q} }
\end{equation}
$\mathtt{qtlsolver}$ was able to find a model (with bound 20, in this case) for formula $\eqref{eq:Spikes} \wedge \eqref{eq:SpikesPast} \wedge \eqref{eq:SpikesPastqPer}$ in around 10 minutes.
As mentioned above, one can add constraints at different levels of abstraction.
For example, we can add SMT constraints imposing that the \emph{values} of the clocks (instead of the clock regions) associated with propositions $p$ and $q$ be periodic; this allows us to check that formula $\eqref{eq:Spikes} \wedge \eqref{eq:SpikesPast} \wedge \eqref{eq:SpikesPastqPer}$ admits periodic models ($\mathtt{qtlsolver}$ takes around 15 minutes to produce one with bound 20).
Finally, if in Formula \eqref{eq:SpikesPastqPer} we replace $\Gltl{(0,100)}{}$ with $\Gltl{(0,100]}{}$, the behavior becomes strictly aperiodic.
In this case the solver takes around 80 minutes to find a model with bound 30, and in excess of 12 hours to show that, with that bound, no model exists in which $p$ and $q$ are periodic (i.e., that the specification, with the added constraint that the values of the clocks associated with $p$ and $q$ be periodic, is unsatisfiable).

While the results presented above are promising, further research will focus on optimizing the implementation of the solver and on extending the encoding to deal with richer constraints.

The techniques presented in this paper for MITL can be tailored also to other logics. We consider an example here. 
A syntactic fragment of MITL was proposed in \cite{HRS98a}, namely MTL$_{0,\infty}$, where temporal modalities are restricted only to intervals of the form 
$\langle 0,b\rangle$ or $\langle a, \infty)$ (e.g., the MITL formula $\mathbf{F}_{(2,3)}\phi$ is not acceptable).  MTL$_{0,\infty}$ is
complete in the sense that every MITL formula can be transformed
into an equisatisfiable MTL$_{0,\infty}$ formula. However, the transformation may lead to an exponential blow-up, since satisfiability is EXPSPACE-complete for MITL and PSPACE-complete for 
  MTL$_{0,\infty}$. 
In \cite{HRS98a}, MTL$_{0,\infty}$ was shown to be equivalent to a new temporal logic, called Event-Clock Logic (ECL), which is also in PSPACE.
Although our work only concerns MITL (and actually MTL$_{0,\infty}$, which is considered by our translation provided that operator $\mathbf{F}_{\langle a,b]}$ is not primitive for the language), our results can directly be applied for solving the satisfiability of (MTL$_{0,\infty}$ and) ECL as well, by means of the above equivalence of the languages. 
However, an explicit encoding  of ECL into CLTL-oc may be devised, since only a finite number of explicit clocks are enough to capture ECL semantics; this may allow solving satisfiability of both logics (MTL$_{0,\infty}$ and ECL) in PSPACE.

%% file: MITL-to-CLTL.bbl
\begin{thebibliography}{10}
\providecommand{\bibitemdeclare}[2]{}
\providecommand{\surnamestart}{}
\providecommand{\surnameend}{}
\providecommand{\urlprefix}{Available at }
\providecommand{\url}[1]{\texttt{#1}}
\providecommand{\href}[2]{\texttt{#2}}
\providecommand{\urlalt}[2]{\href{#1}{#2}}
\providecommand{\doi}[1]{doi:\urlalt{http://dx.doi.org/#1}{#1}}
\providecommand{\bibinfo}[2]{#2}

\bibitemdeclare{misc}{qtlsolver}
\bibitem{qtlsolver}
\emph{\bibinfo{title}{qtlsolver}}.
\newblock \bibinfo{howpublished}{available from
  \href{http://qtlsolver.googlecode.com}{\texttt{qtlsolver.googlecode.com}}}.

\bibitemdeclare{misc}{ae2zot}
\bibitem{ae2zot}
\emph{\bibinfo{title}{Zot: a Bounded Satisfiability Checker}}.
\newblock \bibinfo{howpublished}{available from
  \href{http://zot.googlecode.com}{\texttt{zot.googlecode.com}}}.

\bibitemdeclare{article}{Alur&Dill94}
\bibitem{Alur&Dill94}
\bibinfo{author}{Rajeev \surnamestart Alur\surnameend} \&
  \bibinfo{author}{David~L. \surnamestart Dill\surnameend}
  (\bibinfo{year}{1994}): \emph{\bibinfo{title}{A theory of timed automata}}.
\newblock {\sl \bibinfo{journal}{Theor. Comp. Sci.}}
  \bibinfo{volume}{126}(\bibinfo{number}{2}), pp. \bibinfo{pages}{183--235}.
\newblock \urlprefix\url{http://dx.doi.org/10.1016/0304-3975(94)90010-8}.

\bibitemdeclare{article}{AFH96}
\bibitem{AFH96}
\bibinfo{author}{Rajeev \surnamestart Alur\surnameend}, \bibinfo{author}{Tom{\'
  a}s \surnamestart Feder\surnameend} \& \bibinfo{author}{Thomas~A.
  \surnamestart Henzinger\surnameend} (\bibinfo{year}{1996}):
  \emph{\bibinfo{title}{The Benefits of Relaxing Punctuality}}.
\newblock {\sl \bibinfo{journal}{Journal of the ACM}}
  \bibinfo{volume}{43}(\bibinfo{number}{1}), pp. \bibinfo{pages}{116--146}.
\newblock \urlprefix\url{http://doi.acm.org/10.1145/112600.112613}.

\bibitemdeclare{book}{BK08}
\bibitem{BK08}
\bibinfo{author}{Christel \surnamestart Baier\surnameend} \&
  \bibinfo{author}{Joost-Pieter \surnamestart Katoen\surnameend}
  (\bibinfo{year}{2008}): \emph{\bibinfo{title}{Principles of Model Checking}}.
\newblock \bibinfo{publisher}{MIT Press}.

\bibitemdeclare{incollection}{BY04}
\bibitem{BY04}
\bibinfo{author}{Johan \surnamestart Bengtsson\surnameend} \&
  \bibinfo{author}{Wang \surnamestart Yi\surnameend} (\bibinfo{year}{2004}):
  \emph{\bibinfo{title}{Timed Automata: Semantics, Algorithms and Tools}}.
\newblock In: {\sl \bibinfo{booktitle}{Lect. on Concurrency and Petri Nets}},
  {\sl \bibinfo{series}{LNCS}} \bibinfo{volume}{3098},
  \bibinfo{publisher}{Springer}, pp. \bibinfo{pages}{87--124}.
\newblock \urlprefix\url{http://dx.doi.org/10.1007/978-3-540-27755-2_3}.

\bibitemdeclare{inproceedings}{BFMPRS10}
\bibitem{BFMPRS10}
\bibinfo{author}{Marcello~M. \surnamestart Bersani\surnameend},
  \bibinfo{author}{Achille \surnamestart Frigeri\surnameend},
  \bibinfo{author}{Angelo \surnamestart Morzenti\surnameend},
  \bibinfo{author}{Matteo \surnamestart Pradella\surnameend},
  \bibinfo{author}{Matteo \surnamestart Rossi\surnameend} \&
  \bibinfo{author}{Pierluigi \surnamestart {San Pietro}\surnameend}
  (\bibinfo{year}{2010}): \emph{\bibinfo{title}{Bounded Reachability for
  Temporal Logic over Constraint Systems}}.
\newblock In: {\sl \bibinfo{booktitle}{TIME}}, \bibinfo{publisher}{IEEE
  Computer Society}, pp. \bibinfo{pages}{43--50}.
\newblock \urlprefix\url{http://dx.doi.org/10.1109/TIME.2010.21}.

\bibitemdeclare{misc}{BFMPRS12}
\bibitem{BFMPRS12}
\bibinfo{author}{Marcello~M. \surnamestart Bersani\surnameend},
  \bibinfo{author}{Achille \surnamestart Frigeri\surnameend},
  \bibinfo{author}{Angelo \surnamestart Morzenti\surnameend},
  \bibinfo{author}{Matteo \surnamestart Pradella\surnameend},
  \bibinfo{author}{Matteo \surnamestart Rossi\surnameend} \&
  \bibinfo{author}{Pierluigi \surnamestart {San Pietro}\surnameend}
  (\bibinfo{year}{2012}): \emph{\bibinfo{title}{{CLTL Satisfiability Checking
  without Automata}}}.
\newblock
  \bibinfo{howpublished}{\href{http://arxiv.org/abs/1205.0946}{arXiv:1205.0946%
v1}}.

\bibitemdeclare{incollection}{BFRS11}
\bibitem{BFRS11}
\bibinfo{author}{Marcello~M. \surnamestart Bersani\surnameend},
  \bibinfo{author}{Achille \surnamestart Frigeri\surnameend},
  \bibinfo{author}{Matteo \surnamestart Rossi\surnameend} \&
  \bibinfo{author}{Pierluigi \surnamestart {San Pietro}\surnameend}
  (\bibinfo{year}{2011}): \emph{\bibinfo{title}{Completeness of the Bounded
  Satisfiability Problem for Constraint {LTL}}}.
\newblock In: {\sl \bibinfo{booktitle}{Reachability Problems}}, {\sl
  \bibinfo{series}{LNCS}} \bibinfo{volume}{6945}, pp. \bibinfo{pages}{58--71}.
\newblock \urlprefix\url{http://dx.doi.org/10.1007/978-3-642-24288-5_7}.

\bibitemdeclare{inproceedings}{BRS13b}
\bibitem{BRS13b}
\bibinfo{author}{Marcello~M. \surnamestart Bersani\surnameend},
  \bibinfo{author}{Matteo \surnamestart Rossi\surnameend} \&
  \bibinfo{author}{Pierluigi \surnamestart {San Pietro}\surnameend}
  (\bibinfo{year}{2013}): \emph{\bibinfo{title}{A Tool for Deciding the
  Satisfiability of Continuous-time Metric Temporal Logic}}.
\newblock In: {\sl \bibinfo{booktitle}{Proceedings of the International
  Symposium on Temporal Representation and Reasoning (TIME)}}.
\newblock \bibinfo{note}{To appear}.

\bibitemdeclare{article}{BCM10}
\bibitem{BCM10}
\bibinfo{author}{Patricia \surnamestart Bouyer\surnameend},
  \bibinfo{author}{Fabrice \surnamestart Chevalier\surnameend} \&
  \bibinfo{author}{Nicolas \surnamestart Markey\surnameend}
  (\bibinfo{year}{2010}): \emph{\bibinfo{title}{On the expressiveness of {TPTL}
  and {MTL}}}.
\newblock {\sl \bibinfo{journal}{Information and Computation}}
  \bibinfo{volume}{208}(\bibinfo{number}{2}), pp. \bibinfo{pages}{97 -- 116}.
\newblock \urlprefix\url{http://dx.doi.org/10.1016/j.ic.2009.10.004}.

\bibitemdeclare{article}{DD07}
\bibitem{DD07}
\bibinfo{author}{St\'{e}phane \surnamestart Demri\surnameend} \&
  \bibinfo{author}{Deepak \surnamestart D'Souza\surnameend}
  (\bibinfo{year}{2007}): \emph{\bibinfo{title}{An automata-theoretic approach
  to constraint {LTL}}}.
\newblock {\sl \bibinfo{journal}{Information and Computation}}
  \bibinfo{volume}{205}(\bibinfo{number}{3}), pp. \bibinfo{pages}{380--415}.
\newblock \urlprefix\url{http://dx.doi.org/10.1016/j.ic.2006.09.006}.

\bibitemdeclare{incollection}{Souza&Tabareau04}
\bibitem{Souza&Tabareau04}
\bibinfo{author}{Deepak \surnamestart D'Souza\surnameend} \&
  \bibinfo{author}{Nicolas \surnamestart Tabareau\surnameend}
  (\bibinfo{year}{2004}): \emph{\bibinfo{title}{On Timed Automata with
  Input-Determined Guards}}.
\newblock In: {\sl \bibinfo{booktitle}{Proc. of FORMATS/FTRTFT}}, {\sl
  \bibinfo{series}{LNCS}} \bibinfo{volume}{3253},
  \bibinfo{publisher}{Springer}, pp. \bibinfo{pages}{68--83}.
\newblock \urlprefix\url{http://dx.doi.org/10.1007/978-3-540-30206-3_7}.

\bibitemdeclare{book}{FMMR12}
\bibitem{FMMR12}
\bibinfo{author}{Carlo~A. \surnamestart Furia\surnameend},
  \bibinfo{author}{Dino \surnamestart Mandrioli\surnameend},
  \bibinfo{author}{Angelo \surnamestart Morzenti\surnameend} \&
  \bibinfo{author}{Matteo \surnamestart Rossi\surnameend}
  (\bibinfo{year}{2012}): \emph{\bibinfo{title}{Modeling Time in Computing}}.
\newblock \bibinfo{series}{{EATCS} Monographs in Theoretical Computer Science},
  \bibinfo{publisher}{Springer}.
\newblock \urlprefix\url{http://dx.doi.org/10.1007/978-3-642-32332-4}.

\bibitemdeclare{inproceedings}{HRS98a}
\bibitem{HRS98a}
\bibinfo{author}{Thomas~A. \surnamestart Henzinger\surnameend},
  \bibinfo{author}{Jean~F. \surnamestart Raskin\surnameend} \&
  \bibinfo{author}{Pierre~Y. \surnamestart Schobbens\surnameend}
  (\bibinfo{year}{1998}): \emph{\bibinfo{title}{{The Regular Real-Time
  Languages}}}.
\newblock In: {\sl \bibinfo{booktitle}{Proc. of ICALP'98}}, {\sl
  \bibinfo{series}{LNCS}} \bibinfo{volume}{1343}, pp.
  \bibinfo{pages}{580--591}.
\newblock \urlprefix\url{http://dx.doi.org/10.1007/BFb0055086}.

\bibitemdeclare{article}{HR04}
\bibitem{HR04}
\bibinfo{author}{Yoram \surnamestart Hirshfeld\surnameend} \&
  \bibinfo{author}{Alexander~Moshe \surnamestart Rabinovich\surnameend}
  (\bibinfo{year}{2004}): \emph{\bibinfo{title}{Logics for Real Time:
  Decidability and Complexity}}.
\newblock {\sl \bibinfo{journal}{Fundamenta Informaticae}}
  \bibinfo{volume}{62}(\bibinfo{number}{1}), pp. \bibinfo{pages}{1--28}.

\bibitemdeclare{incollection}{MNP06}
\bibitem{MNP06}
\bibinfo{author}{Oded \surnamestart Maler\surnameend}, \bibinfo{author}{Dejan
  \surnamestart Nickovic\surnameend} \& \bibinfo{author}{Amir \surnamestart
  Pnueli\surnameend} (\bibinfo{year}{2006}): \emph{\bibinfo{title}{From {MITL}
  to Timed Automata}}.
\newblock In: {\sl \bibinfo{booktitle}{Proc. of FORMATS}}, {\sl
  \bibinfo{series}{LNCS}} \bibinfo{volume}{4202}, pp.
  \bibinfo{pages}{274--289}.
\newblock \urlprefix\url{http://dx.doi.org/10.1007/11867340_20}.

\bibitemdeclare{misc}{z3}
\bibitem{z3}
\bibinfo{author}{\surnamestart {Microsoft Research}\surnameend}
  (\bibinfo{year}{2009}): \emph{\bibinfo{title}{Z3: An Efficient {SMT}
  Solver}}.
\newblock \bibinfo{howpublished}{Available at:
  http://research.microsoft.com/en-us/um/redmond/projects/z3/}.

\bibitemdeclare{article}{MS94}
\bibitem{MS94}
\bibinfo{author}{Angelo \surnamestart Morzenti\surnameend} \&
  \bibinfo{author}{Pierluigi \surnamestart {San Pietro}\surnameend}
  (\bibinfo{year}{1994}): \emph{\bibinfo{title}{Object-Oriented Logical
  Specification of Time-Critical Systems}}.
\newblock {\sl \bibinfo{journal}{ACM Transactions on Software Engineering and
  Methodology ({TOSEM})}} \bibinfo{volume}{3}(\bibinfo{number}{1}), pp.
  \bibinfo{pages}{56--98}.
\newblock \urlprefix\url{http://doi.acm.org/10.1145/174634.174636}.

\bibitemdeclare{article}{PMS12}
\bibitem{PMS12}
\bibinfo{author}{Matteo \surnamestart Pradella\surnameend},
  \bibinfo{author}{Angelo \surnamestart Morzenti\surnameend} \&
  \bibinfo{author}{Pierluigi \surnamestart {San Pietro}\surnameend}
  (\bibinfo{year}{2013}): \emph{\bibinfo{title}{Bounded Satisfiability Checking
  of Metric Temporal Logic Specifications}}.
\newblock {\sl \bibinfo{journal}{{ACM} Trans. on Soft. Eng. and Meth.
  ({TOSEM})}}.
\newblock \bibinfo{note}{To appear}.

\end{thebibliography}
